%% file: main.tex
\documentclass{article} % For LaTeX2e
\usepackage{iclr2027_conference,times}

\input{iclr2027/math_commands.tex}

\usepackage{hyperref}
\usepackage{url}
\usepackage{xurl}
\usepackage{graphicx}
\usepackage{float}
\usepackage{placeins}
\usepackage{subcaption}
\usepackage{booktabs}
\usepackage{tabularx}
\usepackage{longtable}
\usepackage{enumitem}
\usepackage[T1]{fontenc}
\usepackage{textcomp}
\setlist[itemize]{
  label=\textbullet,
  leftmargin=1.3em,
  labelwidth=0.7em,
  labelsep=0.3em,
  topsep=2pt,
  itemsep=1pt,
  parsep=0pt,
  partopsep=0pt
}

\usepackage{listings}
\usepackage{tcolorbox}
\tcbuselibrary{listings,breakable,skins}
\definecolor{diffadd}{HTML}{E6FFEC}
\definecolor{diffdelete}{HTML}{FFEBE9}
\definecolor{diffrule}{HTML}{D7E3E9}
\makeatletter
\lst@Key{linebackgroundcolor}{}{\def\DiffLineColor{#1}}
\lst@AddToHook{EveryPar}{%
  \ifx\DiffLineColor\@empty\else
    \rlap{\begingroup\DiffLineColor
      \kern-5pt\rule[-\dp\strutbox]{\dimexpr\linewidth+10pt\relax}{\baselineskip}%
    \endgroup}%
  \fi}
\newcommand{\DiffWrappedLine}{%
  \ifx\DiffLineColor\@empty\else
    \rlap{\begingroup\DiffLineColor
      \kern-\lst@breakcurrindent%
      \rule[-\dp\strutbox]{\dimexpr\linewidth+10pt\relax}{\baselineskip}%
    \endgroup}%
  \fi}
\makeatother
\lstdefinestyle{harnessdiff}{
  basicstyle=\ttfamily\fontsize{8}{9.5}\selectfont,
  columns=fullflexible,
  keepspaces=true,
  breaklines=true,
  breakatwhitespace=true,
  postbreak=\DiffWrappedLine,
  showstringspaces=false,
  aboveskip=5pt,
  belowskip=5pt,
  frame=tb,
  abovecaptionskip=0pt,
  rulecolor=\color{diffrule},
  framerule=.35pt,
  framesep=5pt,
  xleftmargin=5pt,
  xrightmargin=5pt,
  numbers=none,
  firstnumber=1,
  numberblanklines=true,
  tabsize=4,
  upquote=true
}
\ExplSyntaxOn
\cs_generate_variant:Nn \clist_if_in:nnTF { neTF }
\NewDocumentCommand{\DiffBackground}{mm}{
  \clist_if_in:neTF {#1} {\int_eval:n{\value{lstnumber}}}
    {\color{diffadd}}
    {\clist_if_in:neTF {#2} {\int_eval:n{\value{lstnumber}}}
      {\color{diffdelete}} {\color{white}}}
}
\ExplSyntaxOff
\title{\boxbox{}: Recursive Harness Self-Improvement for Long-Running Performance Optimization}

\newcommand{\preprintauthorblock}{%
\begin{minipage}{\textwidth}
\centering
\setlength{\parskip}{0pt}
{\normalfont\fontsize{11}{15}\selectfont
Yuhao~Tan\textsuperscript{1,\textdagger}\quad
Hanyun~Jiang\textsuperscript{2,\textdagger}\quad
Fangkai~Yang\textsuperscript{4,*}\quad
Yuan~Yao\textsuperscript{1}\\
Taolue~Chen\textsuperscript{3}\quad
Bo~Qiao\textsuperscript{4}\quad
Yu~Kang\textsuperscript{4}\quad
Lu~Wang\textsuperscript{4}\quad
Pu~Zhao\textsuperscript{4}\\
Xiaoxing~Ma\textsuperscript{1,*}\quad
Qingwei~Lin\textsuperscript{4}\quad
Saravan~Rajmohan\textsuperscript{4}\quad
Qi~Zhang\textsuperscript{4}\par}
\vspace{7pt}
{\normalfont\fontsize{9}{12}\selectfont
\textsuperscript{1}Nanjing University\quad
\textsuperscript{2}Zhejiang University\\
\textsuperscript{3}Birkbeck, University of London\quad
\textsuperscript{4}Microsoft\par}
\vspace{5pt}
{\normalfont\small
\texttt{yhtan@smail.nju.edu.cn},\quad
\texttt{fangkaiyang@microsoft.com}\par
\vspace{5pt}
\textsuperscript{\textdagger}Work done at Microsoft.\quad
\textsuperscript{*}Corresponding authors.\par
\vspace{6pt}
Project page: \url{http://aka.ms/boxbox}\par}
\end{minipage}%
}

\author{\preprintauthorblock}

\makeatletter
\renewcommand{\@maketitle}{%
  \vbox{%
    \hsize\textwidth
    \parindent0pt
    \parskip0pt
    {\LARGE\scshape\@title\par}
    \vskip10pt
    \nointerlineskip
    \noindent\@author\par
    \vskip12pt
  }%
}
\makeatother

\hypersetup{
  hidelinks,
  pdftitle={BOXBOX: Recursive Harness Self-Improvement for Long-Running Performance Optimization},
  pdfauthor={Yuhao Tan, Hanyun Jiang, Fangkai Yang, Yuan Yao, Taolue Chen, Bo Qiao, Yu Kang, Lu Wang, Pu Zhao, Xiaoxing Ma, Qingwei Lin, Saravan Rajmohan, Qi Zhang}
}

\newcommand{\boxbox}{\textsc{BoxBox}}
\newcommand{\he}{\textsc{HE}}
\newcommand{\hs}{\textsc{HS}}
\newcommand{\frozen}{\textsc{Frozen}}
\newcommand{\perfrace}{\textsc{PerfRace}}
\newcommand{\speedupat}[1]{\ensuremath{\mathrm{Speedup@}#1}}

\iclrfinalcopy
\begin{document}

\maketitle
% The preprint has no conference publication or review banner.
\fancyhead{}
\renewcommand{\headrulewidth}{0pt}

%%%%%%%%%%%%%%%%%%%%%%%%%%%%%%%%%%%

\input{src/1_abstract}

% Keep the lead figure compact so surrounding text can paginate naturally.
\begingroup
\setlength{\intextsep}{6pt}
\begin{figure}[H]
  \centering
  \captionsetup{skip=6pt}
  % Source: artifacts/plot_boxbox_main_results.py.
  \begin{subfigure}[t]{0.49\textwidth}
    \centering
    \includegraphics[width=\linewidth]{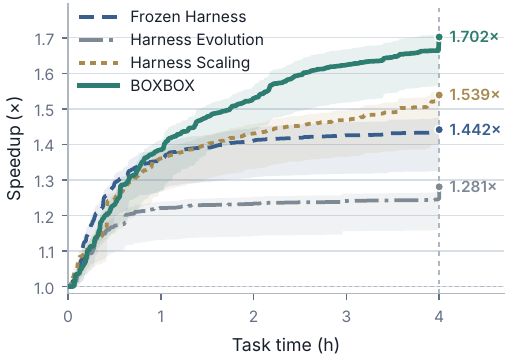}
    \caption{Task performance}
    \label{fig:boxbox-main-results-time}
  \end{subfigure}
  \hfill
  \begin{subfigure}[t]{0.49\textwidth}
    \centering
    \includegraphics[width=\linewidth]{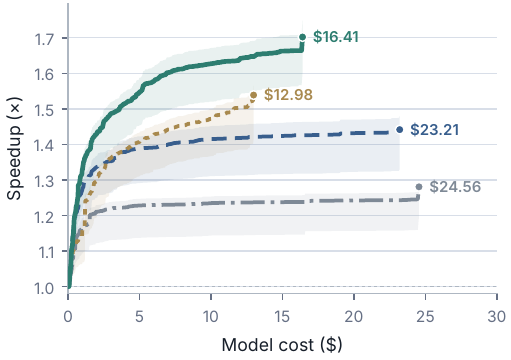}
    \caption{Cost efficiency}
    \label{fig:boxbox-main-results-cost}
  \end{subfigure}
  \caption{Performance of \boxbox{} and baselines on 50 \perfrace{} test tasks,
  with three runs per method and task.
  Curves show aggregate best-so-far correct speedup against
  task time \subref{fig:boxbox-main-results-time} and mean cumulative
  model cost per run \subref{fig:boxbox-main-results-cost},
  sampled at matching task times, with shaded bands indicating
  95\% fixed-task run-resampling intervals.
  \boxbox{} achieves higher speedup with greater time and cost efficiency than the baselines.}
  \label{fig:boxbox-main-results}
\end{figure}
\endgroup

%%%%%%%%%%%%%%%%%%%%%%%%%%%%%%%%%%%%%%%%%%%%%%%%%%%%%%%%%%%%%%%%%%%%%%%

\input{src/2_introduction}
\input{src/3_relatedwork}
\input{src/4_perfrace}
\input{src/5_boxbox}
\input{src/6_experiment}
\input{src/7_conclusion}

% ===============================================================================
\clearpage
\bibliography{iclr2027_conference}
\bibliographystyle{iclr2027_conference}

\newpage
\appendix
\input{src/8a_experiment_appendix.tex}
\clearpage
\input{src/8b_perfrace_appendix.tex}
\clearpage
\input{src/8c_boxbox_appendix.tex}
\clearpage
\input{src/8d_prompt_and_skill.tex}

\end{document}

%% file: iclr2027/math_commands.tex
\usepackage{amsmath,amsfonts,bm}

\def\eqref#1{equation~\ref{#1}}
\def\1{\bm{1}}

\DeclareMathAlphabet{\mathsfit}{\encodingdefault}{\sfdefault}{m}{sl}
\SetMathAlphabet{\mathsfit}{bold}{\encodingdefault}{\sfdefault}{bx}{n}

%% file: src/1_abstract.tex
%   Harness 会影响 agent performance
%                       ↓
%   现有 harness optimization 通常通过专门的优化过程，
%   从任务轨迹和，对照明确的 reference 学习并改进 harness, 
%   用预定义好的 success rate 或者 end-state outcome 进行评估
%                       ↓
%   但许多真实任务是 open-ended的
%                       ↓
%   提出 PerfRace，评价一个 run 内持续产生的进展
%                       ↓
%   我们进一步研究 harness optimization 能不能作为一种在长程任务中
%   由 agent 控制的、递归的能力
%                       ↓
%   提出 BOXBOX (F1 赛车里面汽车进站维修的行为，我们用它来比喻 agent 可以一边跑一边“换轮子”)：
%   agent-triggered、in-place、runtime harness optimization

\begin{abstract}
Automated harness optimization improves agents by learning from trajectories and task feedback. Existing offline methods use reference answers to optimize harnesses for reuse on held-out tasks. However, many real-world tasks are open-ended, while long runs expose new difficulties that offline-optimized harnesses may not address. To study harness optimization in these settings, we introduce \perfrace{}, a benchmark for long-running performance optimization of software projects, and investigate whether agents can use task-specific experience to recursively improve their harnesses during an ongoing task. To this end, we introduce \boxbox{}, which enables agent-triggered harness revision and kernel-managed hot reloading while preserving execution context. We evaluate \boxbox{} on \perfrace{} under matched test-time budgets, comparing it with Frozen Harness, offline Harness Evolution, and Harness Scaling. \boxbox{} achieves higher speedup with more efficient use of task time and model cost, reaching a \speedupat{4\mathrm{h}} of \(1.70\times\), compared with \(1.54\times\) for the strongest baseline.
% Code is available at \url{https://anonymous.4open.science/r/BoxBox-1C35/}.
\end{abstract}

%% file: src/2_introduction.tex
\section{Introduction}
\label{sec:introduction}

%   harness: governs how the agent can interact with the environment
%   harness design can affect agent performance
%   -> 
%   motivates the need for automated harness optimization
%   -> 
%   existing offline harness optimization methods use reference-based 
%   evaluation feedback to diagnose failures and guide harness revisions
%   for reuse on unseen tasks (同类型任务), align with the setting of current benchmarks
%   ->
%   yet real-world tasks often 
%       1. make it difficult to identify useful harness revisions
%       2. as a task progresses, interaction reveals task-specific
%          difficulties that harness revisions learned from other tasks
%          may not adequately address (TTHE, Rethinking),  
%           long-running 会持续遇到很多没有在 held-in 学习过的问题
%   -> 
%   we introduce PerfRace, 有一套 real-wolrd xxx
%   -> 
%   on PerfRace, the offline HE checkpoint with the highest validation
%   score still underperforms Frozen Harness on test tasks
%   -> 
%   we therefore investigate online, task-specific harness optimization
%   ->
%   existing methods use an outer loop to allocate budget and control
%   harness updates, limiting responses to emerging difficulties:
%       (1) the agent observes task progress but cannot decide how to
%           allocate its remaining budget between task execution and
%           harness revision (double-confirm, confilit)
%       (2) updates wait until an attempt or execution segment ends,
%           delaying their use in the ongoing task
%   -> 
%   We introduce BOXBOX, agent-trigger, kernel management
%   -> 
%   Evaluation

Large language model (LLM) agents interact with their environment through harness that determines both the available actions and the observations they receive~\citep{yang2024sweagent,merrill2026terminalbench,lin2026agentic}.
Recent studies have shown that harness design can substantially affect agent performance and behavior, even when the underlying model is fixed ~\citep{lin2026agentic,wang2026harnessevolution}.
These findings have motivated automatic optimization of agent harness, in which an
optimizer analyzes agent trajectories and evaluation feedback, proposes modifications
to the harness, and evaluates the resulting agent on benchmark tasks.
%~\citep{lin2026agentic,wang2026harnessevolution}.

%%%%%%%%%%%%%%%%%%%%%%%%%%%%%%

Existing offline harness optimization methods~\citep{park2026autosaddler,zhang2026selfharness,lin2026agentic,wang2026harnessevolution} often use reference answers or task-completion feedback to diagnose failures, and revise harnesses to improve performance on held-out tasks. This setup aligns with existing agent benchmarks that evaluate whether an agent completes a specified task~\citep{jimenez2024swebench,merrill2026terminalbench,froger2026gaia2}.

However, many real-world tasks are open-ended and long-running, making end-state evaluations alone insufficient to capture how agents progress throughout these runs.
Moreover, this setting poses two challenges for offline harness optimization. First, open-ended tasks make it difficult to identify useful harness revisions, as they may lack clear criteria for diagnosing what needs improvement. Prior evaluations show that harness evolution does not reliably improve performance without unit-test feedback~\citep{wang2026harnessevolution}. Second, continued execution exposes new task-specific difficulties that may limit the generalization of harness revisions learned from other tasks~\citep{nie2026tthe,wang2026harnessevolution}.

%%%%%%%%%%%%%%%%%%%%%%%%%%%%%%%%

% \tl{I think a logic could be: we identify a concrete class of tasks on which typical methods fail; we introduce a new method; we curate benchmark to evaluate.}
% \tyh{Agreed. We now follow this logic: identify the limitations of offline harness optimization on open-ended performance tasks, then introduce \boxbox{} for task-specific harness improvement during ongoing execution. \perfrace{} supplies the real-world tasks for this study.}

To study harness optimization in these settings, we introduce \perfrace{}, a benchmark for optimizing performance of real-world software projects. Our experiments show that offline Harness Evolution (\he{}) underperforms Frozen Harness on \perfrace{} (Figure~\ref{fig:boxbox-main-results}), with no consistent improvement during training or validation (Appendix~\ref{app:experiments-he}), consistent with prior findings of unreliable gains and limited generalization~\citep{wang2026harnessevolution}.

% \tyh{Cite results, Rethinking figure1 also confirms us.}

Nevertheless, the current task's trajectory can reveal difficulties encountered during execution and provide task-specific evidence for harness revision~\citep{nie2026tthe}. We therefore investigate whether an agent can use this experience to improve its own harness during the ongoing task, making harness optimization an online, task-specific process.
Existing methods allocate budget between task execution and harness optimization through an outer loop, with harness updates scheduled by predefined triggers or a separate meta-agent~\citep{wang2026harnessevolution,liu2026evox,zhang2026agenticevolution}. However, external control limits the task-solving agent's ability to respond to emerging difficulties. First, despite observing task progress, the agent cannot decide whether to spend its remaining budget on task execution or harness revision. Second, updates occur between task attempts or execution segments, delaying their use in the ongoing task.

To address these limitations, we introduce \boxbox{}\footnote{Named after the F1 radio call ``box, box,'' requesting a pit stop for repairs and adjustments during the race.} to support agent-triggered revision of the full executable harness through kernel-managed hot reloading.
When requested, the kernel pauses task execution and launches an agent review session to diagnose difficulties from the trajectory and revise the harness.
After validation, it resumes the task with preserved execution context under the revised harness, which governs both subsequent execution and future reviews, enabling recursive harness improvement.
We evaluate \boxbox{} on \perfrace{} against Frozen Harness, offline Harness Evolution and externally controlled Harness Scaling under matched task budgets, comparing best-so-far correct speedup and model cost. \boxbox{} achieves a \speedupat{4\mathrm{h}} of \(1.70\times\), compared with \(1.44\times\) for Frozen Harness, \(1.28\times\) for Harness Evolution, and \(1.54\times\) for Harness Scaling.

Our main contributions are summarized as follows:
\begin{itemize}
    \item We introduce \perfrace{}, a benchmark for evaluating agents on open-ended performance optimization tasks, tracking correctness-preserving improvements to real-world software throughout task execution.
    \item We present \boxbox{}, which combines agent-triggered, full-harness revision from task-specific experience with kernel-managed hot reloading of the runtime while preserving execution context.
    \item Our experiments show that \boxbox{} achieves higher best-so-far correct speedup with greater time and cost efficiency than the evaluated baselines, supporting harness optimization as part of an agent's own capabilities.
\end{itemize}

%% file: src/3_relatedwork.tex
\section{Related Work}
\label{sec:related-work}

\textbf{Automated harness optimization.} Prior work has extensively studied the automated optimization of prompts for LLM-based systems~\citep{zhou2023ape,pryzant2023protegi,yuksekgonul2025textgrad,khattab2024dspy,agrawal2025gepa}. Subsequent work extends automated optimization to agents, using execution trajectories and task feedback to improve their memory, skills, and workflows~\citep{zhao2024expel,zhang2025ace,ni2026trace2skill,hu2024adas,zhang2024aflow}. Meta-Harness and AHE further extend automated optimization to executable harness code, using trajectories and evaluation feedback to guide iterative revisions~\citep{lee2026metaharness,lin2026agentic}.

Building on this setting, offline methods revise harnesses using reference answers or task-completion feedback from training tasks, aiming to improve held-out performance~\citep{park2026autosaddler,zhang2026selfharness,lin2026agentic}. However, an evaluation finds unreliable gains without unit-test feedback and limited transfer from reusable harnesses, but reports gains from Harness Scaling, which alternates task attempts and harness revisions at test time~\citep{wang2026harnessevolution}.

Following this task-specific direction, JIT-Agent synthesizes a tailored harness for each task before execution begins~\citep{zhang2026jita}. TTHE further explores harness optimization without reference answers, using trajectories and execution-derived feedback to revise executable harnesses on unlabeled test batches~\citep{nie2026tthe}. Continual Harness moves harness optimization into ongoing task execution through an outer loop that periodically invokes a refiner while preserving the task state~\citep{karten2026continualharness}. In \boxbox{}, the task-solving agent decides when to allocate its remaining budget to full-harness revision, while a separate kernel manages runtime hot reloading and preserves execution context.

\textbf{Long-Horizon and Open-Ended Agent Evaluation.} Existing agent benchmarks cover diverse capabilities but commonly evaluate tasks with explicit end states, reporting task success, pass rates, or final artifact quality~\citep{mialon2023gaia, liu2024agentbench, jimenez2024swebench, xie2024osworld, merrill2026terminalbench, sun2026agentslastexam}. Recent iterative optimization benchmarks evaluate solution quality but still use reference implementations, human performance, or reference outputs for task construction or scoring~\citep{chan2025mlebench,lyu2026mlsbench,imajuku2025alebench,xu2026autolab,mang2025frontiercs,zhu2026edgebench}. Closer to our setting, Frontier-Eng and FrontierSWE support iterative optimization of engineering designs and software projects, with tasks curated from established problems and expert contributions~\citep{chi2026frontiereng,chu2026frontierswe}.

In contrast, \perfrace{} constructs long-running performance optimization tasks from real-world software projects and workloads without requiring reference optimizations. Its construction pipeline provides a scalable way to continually source new tasks from public software projects.

\textbf{Benchmarks for Performance Optimization.} \citet{shypula2024learning} introduce PIE for evaluating LLMs' ability to make performance-improving edits to single-file competitive programming solutions while preserving functional correctness. Subsequent benchmarks broaden the optimization domains to GPU kernels, general-purpose numerical programs, and algorithmic redesign, while retaining function- or file-level optimization tasks~\citep{du2024mercury,huang2024effibench,liu2024evalperf,waghjale2024ecco,ouyang2025kernelbench,press2025algotune}.

GSO and SWE-Perf extend evaluation to repository-level performance optimization, deriving tasks and human reference optimizations from GitHub commits and pull requests~\citep{shetty2025gso,he2025sweperf}. SWE-fficiency and FormulaCode further incorporate a targeted workload per task and multiple community-maintained workloads, respectively~\citep{ma2025swefficiency,sehgal2026formulacode}.

These repository-level benchmarks construct tasks around a targeted performance test, whereas real performance engineering starts from an end-to-end workload and discovers bottlenecks through profiling. \perfrace{} adopts this workload-first formulation, leaving agents to identify and implement correctness-preserving improvements without a known target patch.

%% file: src/4_perfrace.tex
\section{\perfrace{}}
\label{sec:perfrace}

\perfrace{} benchmarks real-world software performance optimization with three design principles: (1) \emph{Live}: projects are continually sourced from public GitHub repositories; (2) \emph{Open-ended}: tasks use project workloads without reference optimizations or known target patches; and (3) \emph{Long horizon}: agents iteratively optimize code and choose when to submit for feedback within a fixed budget.

%%%%%%%%%%%%%%%%%%%%%%%%%%%%%

\subsection{Task Formulation}
\label{sec:perfrace-formulation}

Each \perfrace{} task provides the agent with an editable project workspace,
a fixed workload that the agent can run locally, and access to an external Judge.
For official evaluation, the Judge uses a separate copy of this workload
and additional held-out correctness cases.

During the evaluation loop, the agent iteratively modifies the project and submits changes to the Judge for feedback. Let \(t_i\) denote the evaluation start time of submission \(i\) and \(s_i\) its score, with \(s_i=0\) for incorrect submissions. Scores are attributed retrospectively to evaluation start times; verification time also counts toward the task budget. We include the unmodified project as a baseline submission with \(t_0=0\) and \(s_0=1\). We define the best-so-far correct speedup as \(S(t)=\max_{i:t_i\leq t}s_i\), which the agent aims to maximize within a fixed task budget \(B\).

\begin{figure}[t]
  \centering
  \includegraphics[width=0.95\textwidth]{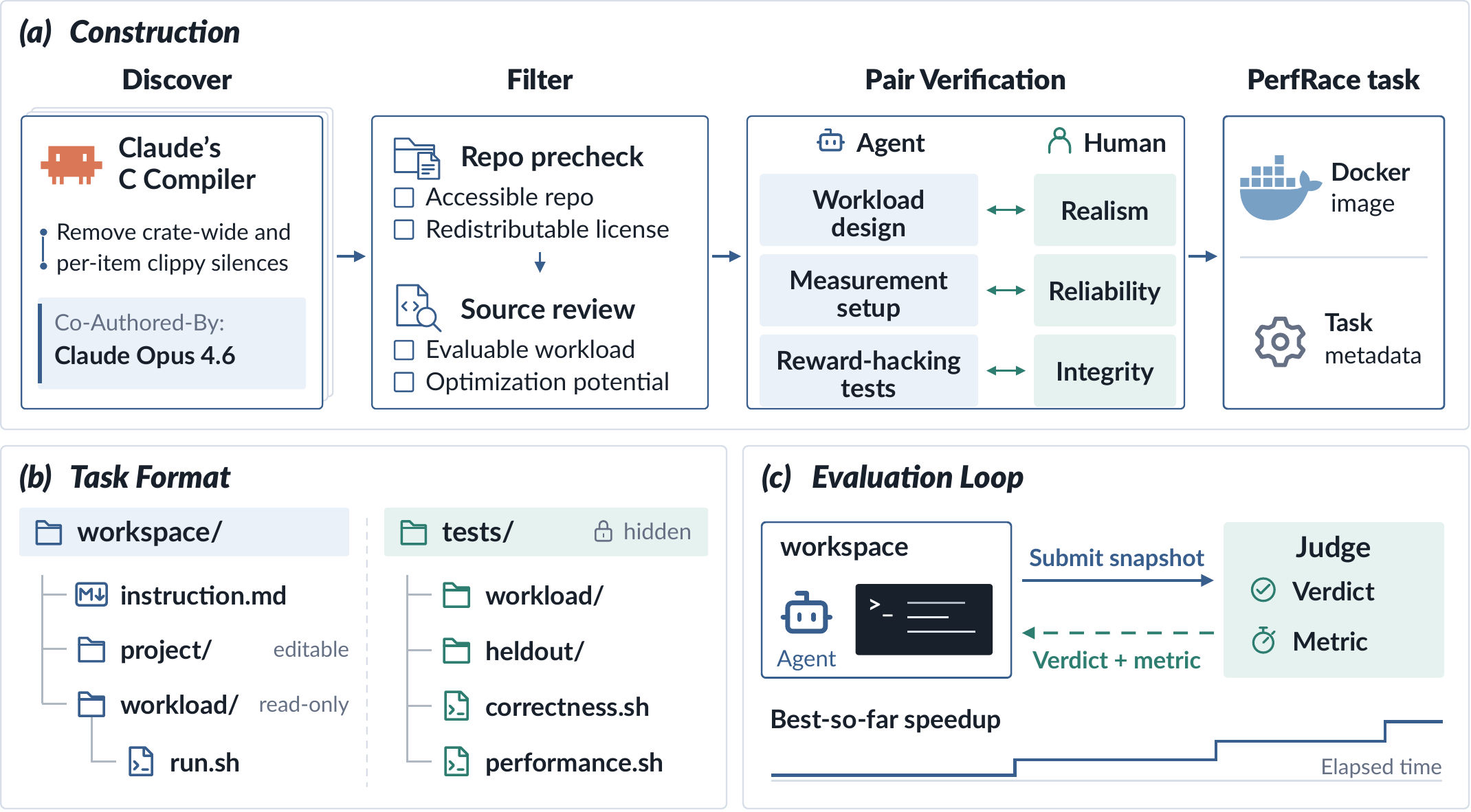}
  \caption{Overview of \perfrace{}.
  (a) Construction discovers projects from GitHub coding-agent activity, applies
  a two-stage Filter comprising a deterministic precheck and source-based
  assessment, and uses Agent--Human Pair Verification to produce a \perfrace{}
  task.
  (b) Task format separates the agent workspace from hidden evaluation files.
  (c) The evaluation loop repeatedly submits workspace snapshots to the
  Judge and tracks best-so-far correct speedup.}
  \label{fig:perfrace-overview}
\end{figure}

%%%%%%%%%%%%%%%%%%%%%%%%%%%%%%%%%%%%%%%%%%%%%%%%%%%%%%%%%%%%
\subsection{Construction}
\label{sec:perfrace-construction}

We construct \perfrace{} by identifying real projects with optimization potential
and building reliable tasks around realistic workloads (Figure~\ref{fig:perfrace-overview}(a)).
Each \perfrace{} task adopts the Harbor Task Format~\citep{merrill2026terminalbench}, as illustrated in Figure~\ref{fig:perfrace-overview}(b).

\textbf{Discovery.} To identify less-explored optimization opportunities beyond mature, heavily maintained repositories, we search GitHub coding-agent activity for candidate projects. Specifically, we identified 103,475 candidate projects from public GitHub pull requests and branch pushes made by known coding agents between May and August 2026 (Appendix~\ref{app:perfrace-curation-statistics}).

\textbf{Filter.} We filter candidate projects in two stages. First, deterministic prechecks verify repository and source-revision accessibility, license redistributability, and repository size and popularity, retaining 7,257 candidates. We then use GPT-5.6-Luna to assess each candidate's setup process, evaluable workload, runtime feasibility, and optimization potential from its source code. Of these candidates, 239 pass this assessment and proceed to Pair Verification (Appendix~\ref{app:perfrace-curation-statistics}).

\textbf{Pair Verification.}
We use agent--human Pair Verification to construct runnable tasks and assess their quality against three criteria (Appendix~\ref{app:perfrace-pair-verification}):
\begin{itemize}
  \item \textbf{Realism.} The workload and metric represent meaningful project use.
  \item \textbf{Reliability.} Repeated equivalent runs support the same correctness and
  performance conclusions.
  \item \textbf{Integrity.} The score reflects genuine task completion rather than an
  invalid or shortcut solution.
\end{itemize}

Specifically, we first let the agent derive workload inputs and scale from real project use and propose correctness checks, measurement boundaries, and performance metrics for human review.
Second, the agent configures warmup and runs repeated tests to detect flakiness and distinguish genuine speedups from noise; the reviewer uses these results to approve or revise the measurement protocol.
Finally, the agent conducts red-team tests against reward hacking using invalid, incomplete, hard-coded, and work-skipping solutions; the reviewer requests verifier fixes as needed before approving the task for packaging and release.
Eventually, we curated 100 tasks, each requiring an average of 9.1 hours of
curation, 12 agent--human interactions, and \$11.35 in agent cost; detailed
statistics are provided in Appendix~\ref{app:perfrace-curation-statistics}.

\subsection{Benchmark Composition}
\label{sec:perfrace-composition}

\begin{figure}[H]
  \centering
  \includegraphics[width=\textwidth]{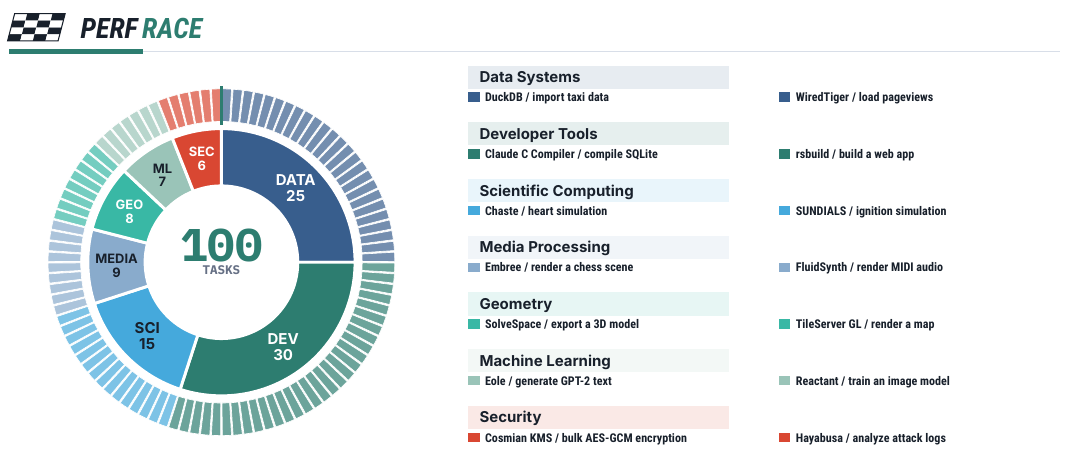}
  \caption{Composition of \perfrace{}: seven workload domains (inner ring), 100 tasks (outer ring), and representative project--workload pairs (right).}
  \label{fig:perfrace-composition}
\end{figure}

The resulting benchmark contains 100 tasks across seven domains (Figure~\ref{fig:perfrace-composition}): \emph{developer tools} (30 tasks) compiles, builds and validates software; \emph{data systems} (25 tasks) ingests, indexes and queries data; \emph{scientific computing} (15 tasks) runs simulations and numerical solvers; \emph{media processing} (9 tasks) renders graphics and audio; \emph{geometry} (8 tasks) processes maps and 3D models; \emph{machine learning} (7 tasks) trains models and runs inference; and \emph{security} (6 tasks) encrypts data and analyzes attack logs.
Each task pairs a real project with a fixed workload; for example, one task optimizes CCC while it compiles SQLite. Appendix~\ref{app:perfrace-characterization} provides a detailed characterization of the tasks.

%%%%%%%%%%%%%%%%%%%%%%%%%%%%%%%%%%%%%%%%%%%%%%%%%%%%%%%%%%%%%%%%%%%

\subsection{Evaluation}
\label{sec:perfrace-evaluation}

In the evaluation loop, the agent chooses when to submit snapshots for correctness and performance feedback from the Judge (Figure~\ref{fig:perfrace-overview}(c)). The score \(s_i\) is the speedup over the unmodified project for correct submissions and zero otherwise. Runs without a correctness-preserving improvement receive a score of \(1\times\). During the final five minutes, the agent is prompted to end optimization and leave the project in a coherent state for the Judge's final evaluation.

Prior studies show that agents can inflate scores by tampering with evaluation code or accessing reference outputs~\citep{wang2026benchjack,metr2025rewardhacking}. We therefore use \emph{shadow verification}, evaluating project snapshots in separate containers while keeping verifier code and held-out test data out of the agent's workspace.
Time spent in shadow verification counts toward \(B\).
Appendix~\ref{app:perfrace-evaluation} provides further details on shadow verification.

%% file: src/5_boxbox.tex
\section{\boxbox{}}
\label{sec:boxbox}

We now present \boxbox{} for agent-triggered harness revision during long-running tasks. The following subsections present the problem formulation, explain the kernel-managed harness update mechanism, and describe a simple self-review strategy for diagnosis and revision.

\subsection{Problem Formulation}
\label{sec:boxbox-formulation}

Consider a fixed language model $\mathcal{M}$ and an executable harness that may change during a task. Let $H_k$, $e_k$, and $\tau_k$ denote the active harness, environment state, and accumulated trajectory at interaction step $k$, respectively~\citep{yao2023react}. Task execution follows
\begin{equation}
  \begin{aligned}
    a_k &\sim p_{\mathcal{M},H_k}(\cdot\mid\tau_k),
    \qquad a_k\in\mathcal{A}(H_k),\\
    (e_{k+1},o_{k+1}) &\leftarrow \operatorname{Execute}_{H_k}(e_k,a_k),\\
    \tau_{k+1} &= \tau_k\oplus(a_k,o_{k+1}),\\
    H_{k+1} &=
    \begin{cases}
      H'_k, & \text{if a revision is activated},\\
      H_k, & \text{otherwise}.
    \end{cases}
  \end{aligned}
  \label{eq:boxbox-interaction}
\end{equation}
Here, $a_k$ is a tool call from the action space $\mathcal{A}(H_k)$ exposed by the active harness, and $p_{\mathcal{M},H_k}$ is its distribution induced by $H_k$ constructing model inputs from $\tau_k$ and invoking $\mathcal{M}$. $\operatorname{Execute}_{H_k}$ resolves and executes $a_k$ through the harness's tool implementation, updates the environment state, and processes the resulting feedback into $o_{k+1}$; $\oplus$ appends the action--observation pair to the history.

When a revision is attempted, the agent revises $H_k$ based on the interaction history $\tau_{k+1}$, producing a candidate $H'_k$. If activated, it becomes the active harness for the next interaction; otherwise, execution continues under $H_k$. Subsequent action selection and execution therefore use $H_{k+1}$.

% The timing and implementation of harness revisions are determined by the optimization method.

\subsection{Kernel-Managed Harness Updates}
\label{sec:boxbox-mechanism}

\boxbox{} treats the harness as a task-specific implementation, allowing a revision to be tailored to the requirements of the ongoing task. To support such adaptations, \boxbox{} exposes the full executable harness, allowing the agent to translate task-specific needs into direct changes to tools, control flow, and state management. Let $\mathcal{H}$ denote the set of harness implementations compatible with the execution interface. Methods with predefined editable components restrict candidate revisions to a subset $\mathcal{H}_{\mathrm{bounded}}\subseteq\mathcal{H}$~\citep{zhang2026selfharness,lin2026agentic,park2026autosaddler}, whereas \boxbox{} allows revisions $H'_k\in\mathcal{H}$.

In \boxbox{}, the harness is the editable agent implementation, while the runtime is its running version that controls the agent's execution and maintains its state. These revisions change the agent's execution code, so a faulty update can interrupt the task and make it difficult for the agent to recover on its own. \boxbox{} therefore introduces a separate kernel that manages runtime hot reloading and can safely retain or restore the previous runtime if an update fails validation or activation.

Specifically, we register \texttt{boxbox()} as a tool that yields control to the kernel, with $\texttt{boxbox()}\in\mathcal{A}(H_k)$. When the agent invokes \texttt{boxbox()}, the kernel suspends task execution and preserves the agent's execution context, including the accumulated trajectory, terminal session, and workspace state. It then launches a review session for the agent to examine the trajectory, revise the harness as needed, and commit any changes. After the review session concludes, the kernel validates any new commits, loads the revised runtime on success, and otherwise retains the current runtime. During activation, the kernel transfers the preserved execution context to the new runtime and reattaches the existing terminal session. On successful activation, $H_{k+1}=H'_k$, and subsequent interactions follow Eqn.~\ref{eq:boxbox-interaction}. Finally, the kernel returns control to the agent, allowing it to resume the ongoing task.

Beyond agent-initiated updates, runtime faults also trigger kernel-managed recovery and review based on the recorded fault evidence. Appendix~\ref{app:boxbox} provides further implementation details.

\begin{figure*}[t]
  \centering
  \includegraphics[width=0.9\textwidth]{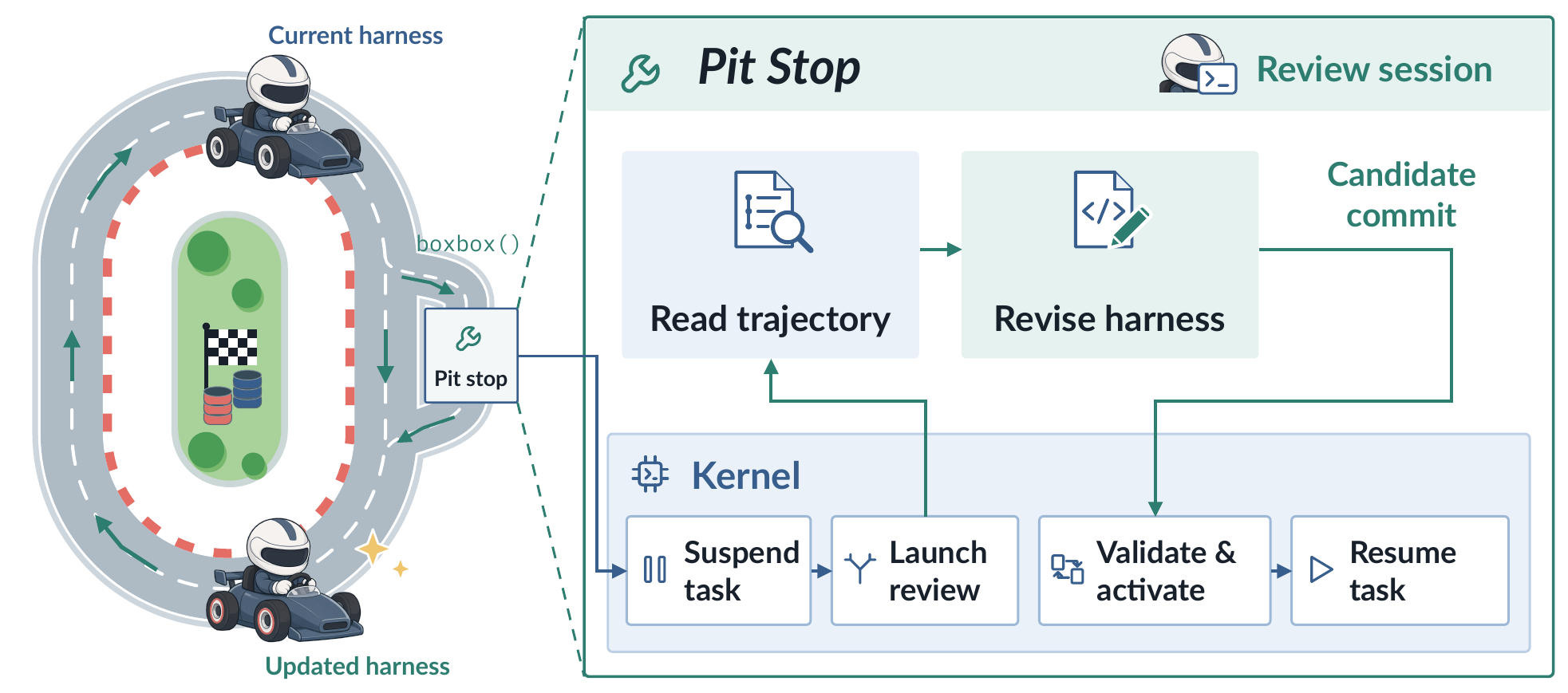}
  \caption{\textbf{\boxbox{} mechanism.}
  The agent requests a review on demand, triggering a trap that transfers control to the kernel. The kernel then launches the review, validates revisions, and hot-reloads the runtime with preserved execution context and rollback if activation fails.}
  \label{fig:boxbox-mechanism}
\end{figure*}

\subsection{Agent-Controlled Harness Optimization}
\label{sec:boxbox-strategy}

Based on this mechanism, \boxbox{} lets the task-solving agent decide when to request a harness review, so it can adapt its harness in response to difficulties encountered during task execution. Unlike prior methods that use an outer loop~\citep{lee2026metaharness,karten2026continualharness,liu2026evox}, the agent itself allocates the shared budget $B$ between project-code optimization and harness revision to maximize $S(B)$.

Within the review session, we provide a default skill (Appendix~\ref{app:boxbox-review-skill}) that guides the agent through a simple self-review strategy based on diagnosis and harness revision:
\begin{itemize}
  \item \textbf{Diagnosis.}
  The agent examines the trajectory for execution failures, repeated work, and successful patterns that could be automated or reused. It uses these observations to identify limitations of the current harness, such as missing capabilities or costly tool interactions.

  \item \textbf{Revision.}
  The agent selects harness revisions based on their expected benefit within the remaining budget, such as adding guidance, improving tools, or changing execution logic. For example, repeated tool calls can be combined into a reusable tool.
\end{itemize}
We track harness revisions as Git commits so the agent can inspect prior changes alongside execution records in subsequent reviews, assess their effects, and decide whether further revisions are needed.

%% file: src/6_experiment.tex
\section{Experiments}
\label{sec:experiments}

\subsection{Experimental Setup}
\label{sec:experiments-setup}

\textbf{Baselines.}
We compare \boxbox{} with three baselines: \textbf{Frozen Harness} (\frozen{}) retains the initial harness throughout task execution. \textbf{Harness Evolution} (\he{}) optimizes the harness offline on training tasks and selects it by validation performance, then holds it fixed during test-task execution. \textbf{Harness Scaling} (\hs{}) uses an external controller to revise the harness between attempts on the same task, using feedback from each completed attempt. Both \he{} and \hs{} are adapted from \citet{wang2026harnessevolution}. Following Meta-Harness~\citep{lee2026metaharness}, all methods start from the same initial harness $H_0$, Terminus-KIRA~\citep{terminuskira2026}, a simple terminal-based harness that executes shell commands and returns their outputs to the model.

\textbf{Evaluation protocol.}
We split the 100 \perfrace{} tasks into 40 training and 10 validation tasks for \he{}, and 50 held-out test tasks for evaluating all methods. Both task execution and harness optimization are performed using GPT-5.6 Luna with high reasoning effort. Each method is evaluated with three independent runs per test task, each sharing a four-hour budget ($B=4\mathrm{h}$) across task execution, harness optimization, and evaluation. Detailed configurations are provided in Appendix~\ref{app:experiments}.

\textbf{Metrics.}
To account for randomness, we summarize task $i$'s best-so-far correct
speedup across $R=3$ runs as
$\bar S_i=\left(\prod_{r=1}^{R}S_i^{(r)}(B)\right)^{1/R}$.
Let $\overline{\mathrm{Cost}}_i$ be its mean model cost in US dollars,
including harness optimization.
Across $N=50$ tasks, we use harmonic mean speedup to limit the influence of
extreme values~\citep{shetty2025gso} and arithmetic mean task-level speedup per dollar:
\begin{equation}
  \speedupat{4\mathrm{h}}=\frac{N}{\sum_{i=1}^{N}\bar S_i^{-1}},
  \qquad
  \mathrm{Speedup/\$}
  =\frac{1}{N}\sum_{i=1}^{N}
    \frac{\bar S_i}{\overline{\mathrm{Cost}}_i}.
  \label{eq:aggregate-speedup}
\end{equation}
Reported mean model cost averages $\overline{\mathrm{Cost}}_i$ across tasks.
Cost accounting and statistical details appear in Appendix~\ref{app:experiments}.

%%%%%%%%%%%%%%%%%%%%%%%%%%%%%%%%%%%%%%%%%%%%%%%%%%%%%%%%%%
\subsection{Main Results}
\label{sec:experiments-main}

We report performance and model cost in Table~\ref{tab:boxbox-main-results} and plot speedup against task time and cumulative model cost in Figure~\ref{fig:boxbox-main-results}. Appendix~\ref{app:experiments-complete-results} provides optimization trajectories for all test tasks.

\input{tables/boxbox-main-results}

\textbf{Offline harness optimization does not reliably improve performance.}
\he{} reaches \(1.281\times\) at four hours, below \frozen{} (Table~\ref{tab:boxbox-main-results}). \he{} struggles to identify consistently effective harness revisions during training, while the revisions selected through validation show limited generalization to long-running test tasks. Appendix~\ref{app:experiments-he} provides detailed results and failure analysis. \he{} costs an average of \$24.56 per test run. Including offline training and validation costs, shared across the 150 test runs, brings this average to at least \$33.73. This additional offline expenditure does not translate into better test performance or greater cost efficiency than \frozen{}.

\textbf{Task-specific harness optimization supports continued improvement.}
Both \hs{} and \boxbox{} improve over \frozen{}, reaching \(1.539\times\) and \(1.702\times\) at four hours, respectively, compared with \(1.442\times\) for \frozen{} (Table~\ref{tab:boxbox-main-results}). Both also reduce mean model cost, to \$12.98 and \$16.43 per run, respectively, compared with \$23.22 for \frozen{}. Figure~\ref{fig:boxbox-main-results} shows the corresponding speedup trajectories against elapsed task time and mean cumulative model cost. Over the second half of the four-hour budget, the \frozen{} curve largely levels off, increasing from \(1.410\times\) to \(1.442\times\), while \hs{} improves from \(1.431\times\) to \(1.539\times\) and \boxbox{} from \(1.530\times\) to \(1.702\times\). These trends show that task-specific harness optimization supports further progress as execution continues.

% Figure 1 source: artifacts/plot_boxbox_main_results.py.

\textbf{Agent-controlled harness optimization improves performance and efficiency.}
Figure~\ref{fig:boxbox-main-results-time} shows that \boxbox{} achieves higher speedups than \hs{} at matched elapsed task time. This result compares \hs{}'s attempt-level external control with \boxbox{}'s finer-grained, agent-triggered revision during task execution. Whereas \hs{} relies on an external controller to initiate revisions, \boxbox{} lets the agent decide when to allocate budget to adapting its harness to the needs of the ongoing task. This allows the agent to address harness limitations as they arise and apply useful revisions directly to the ongoing task. Figure~\ref{fig:boxbox-main-results-cost} further shows higher speedups for \boxbox{} at matched cumulative model costs. Although \hs{} has a lower total model cost, \boxbox{} achieves greater speedup per dollar under our task-level metric. Appendix~\ref{app:experiments-hs} provides further analysis.

\subsection{Ablation Studies and Analysis}
\label{sec:experiments-ablation}

We further conduct ablation studies on 14 randomly sampled test tasks to isolate the contributions of key design choices to the effectiveness and efficiency of harness optimization, then analyze how agents perform harness optimization during task execution.

\textbf{Ablation Studies.}

\input{tables/boxbox-online-ablation}

\textit{Agent-controlled triggering.}
Following the scheduled refinement in Continual Harness~\citep{karten2026continualharness}, the \textit{w/o Agent Triggering} variant schedules reviews every 30 minutes with full-harness editing. It achieves lower speedup at higher model cost than agent-triggered \boxbox{} (Table~\ref{tab:boxbox-online-ablation}). Scheduled reviews can interrupt the process of proposing, implementing, and evaluating code changes. They also consume model calls even when they conclude that no harness revision is needed. In contrast, \boxbox{} lets the agent decide when to allocate budget to harness optimization based on the needs of the ongoing task.

\textit{Full-harness editing.}
Following Self-Harness~\citep{zhang2026selfharness}, the \textit{w/o Full Editing} variant retains agent-controlled triggering but restricts each review to a predefined scope. It achieves lower speedup than \boxbox{} (Table~\ref{tab:boxbox-online-ablation}). Predefined editing boundaries can constrain task-specific adaptations that require changes across components or beyond the permitted scope. Full-harness editing lets the agent decide what to modify and how extensively based on the needs of the ongoing task.

% \textit{Context preservation.}
% The \textit{w/o Context Preservation} variant resets only the task-conversation history after each review, preserving the project workspace, terminal sessions, and running processes.

\textbf{Harness Optimization Analysis.}

\begin{figure}[htbp]
  \centering
  \captionsetup{skip=6pt}
  % Source: artifacts/plot_boxbox_review_analysis.py.
  % Widths follow the panel aspect ratios to keep plot heights aligned.
  \begin{subfigure}[t]{0.2992\textwidth}
    \centering
    \includegraphics[width=\linewidth]{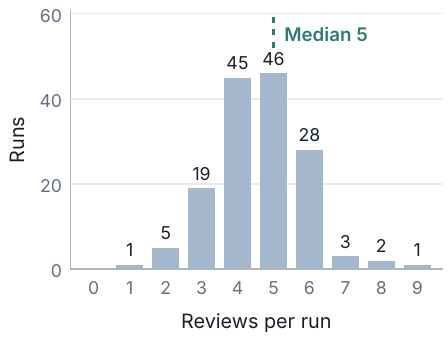}
    \caption{Review frequency}
    \label{fig:boxbox-review-analysis-frequency}
  \end{subfigure}
  \hfill
  \begin{subfigure}[t]{0.2864\textwidth}
    \centering
    \includegraphics[width=\linewidth]{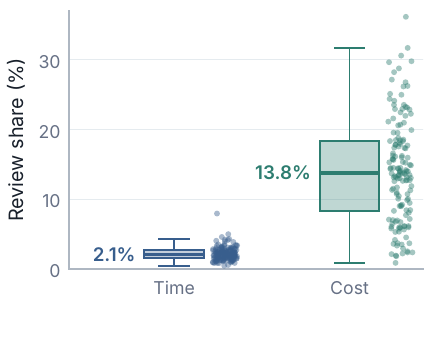}
    \caption{Budget allocation}
    \label{fig:boxbox-review-analysis-budget}
  \end{subfigure}
  \hfill
  \begin{subfigure}[t]{0.3944\textwidth}
    \centering
    \includegraphics[width=\linewidth]{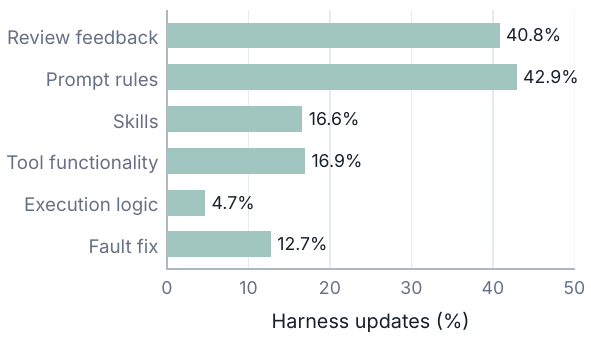}
    \caption{Modification types}
    \label{fig:boxbox-review-analysis-modifications}
  \end{subfigure}
  \caption{\textbf{Harness optimization analysis.}
  \subref{fig:boxbox-review-analysis-frequency} Number of reviews per run across 150 test runs, including agent-triggered and fault-recovery reviews.
  \subref{fig:boxbox-review-analysis-budget} Per-run review time as a share of the four-hour task budget and review cost as a share of total model cost within that budget; dots represent runs, and boxes show the median and interquartile range.
  \subref{fig:boxbox-review-analysis-modifications} Modification types across a collection of 620 applied, nonempty harness updates from \boxbox{} runs.}
  \label{fig:boxbox-review-analysis}
\end{figure}

\textit{Review frequency.}
\boxbox{} invokes a median of five reviews per run, with 92\% of runs containing three to six (Figure~\ref{fig:boxbox-review-analysis-frequency}). These results show that \texttt{boxbox()} provides an effective entry point for harness optimization during task execution, allowing agents to review their harnesses based on accumulated trajectories.

\textit{Budget allocation.}
Agents allocate most of their time budget to task execution. Harness reviews account for a median of 2.1\% of the time budget and 13.8\% of model cost per run (Figure~\ref{fig:boxbox-review-analysis-budget}). Together with \boxbox{}'s higher speedup and lower model cost than \frozen{}, these results suggest that agents can effectively balance task execution and harness optimization within a shared budget.

\textit{Modification scope.}
Harness modifications span prompt rules, skills, tools, execution logic, and fault handling (Figure~\ref{fig:boxbox-review-analysis-modifications}). These changes show that agents use full-harness editing to adapt both their capabilities and execution process. Cases in Appendix~\ref{app:case-study} illustrate how these modifications add capabilities, automate waiting, and revise context management.

The ablations show that agent-controlled triggering and full-harness editing improve performance and cost efficiency, while the analysis shows that agents adapt their capabilities and execution process while allocating most of their time to task execution. Together, these findings support harness optimization as part of the task-solving agent's own capabilities.

% \subsection{Generalization}
% \label{sec:experiments-generalization}

% We use two complementary comparisons to assess whether the benefits of
% \boxbox{} extend beyond the original model and initial harness.
% For harness generalization, we replace Terminus-KIRA with Pi or
% mini-swe-agent while retaining GPT-5.6 Luna.
% For model generalization, we retain Terminus-KIRA and replace GPT-5.6 Luna
% with Gemini-3.8-Flash or Grok-4.6.
% Each configuration compares \frozen{} with \boxbox{}, changing one
% factor at a time rather than evaluating a full model-by-harness grid.
% Table~\ref{tab:boxbox-generalization} compares performance and model cost
% across these configurations.

% \input{tables/boxbox-generalization}

%% file: tables/boxbox-main-results.tex
\begin{table}[t]
  \centering
  \small
  \setlength{\tabcolsep}{5pt}
  \caption{Results on 50 test tasks with three four-hour runs per task, reporting \speedupat{4\mathrm{h}}, mean model cost per run, and mean task-level speedup per dollar.}
  \label{tab:boxbox-main-results}
  \begin{tabular}{lccc}
    \toprule
    Method & \(\speedupat{4\mathrm{h}}\) &
    Mean model cost & Speedup/\$ $\uparrow$ \\
    \midrule
    % HE uses h24 and includes final delivery, consistent with Figure 1.
    \frozen{}  & 1.442 & \$23.22 & 0.2142 \\
    \he{}      & 1.281 & \$24.56 & 0.1243 \\
    \hs{}      & 1.539 & \$12.98 & 0.2872 \\
    \boxbox{}       & \textbf{1.702} & \$16.43 & \textbf{0.4886} \\
    \bottomrule
  \end{tabular}
\end{table}

%% file: tables/boxbox-online-ablation.tex
\begin{table}[t]
  \centering
  \small
  \setlength{\tabcolsep}{6pt}

  \caption{Ablation study of \boxbox{}.
  We examine agent-controlled triggering and full-harness editing
  in terms of task performance (\(\speedupat{4\mathrm{h}}\)), model cost,
  and cost efficiency (Speedup/\$).}
  \label{tab:boxbox-online-ablation}
  \begin{tabular}{lccc}
    \toprule
    Method & \(\speedupat{4\mathrm{h}}\) & Mean model cost & Speedup/\$ \\
    \midrule
    w/o Agent Triggering & 1.508 & \$21.58 & 0.1338 \\
    w/o Full Editing & 1.349 & \$16.07 & 0.1132 \\
    % w/o Context Preservation & 1.460 & \$9.205 & 0.2894 \\
    % w/o Harness Editing & 1.550 & \$26.07 & 0.1673 \\
    Full \boxbox{} & \textbf{1.584} & \$15.07 & \textbf{0.3452} \\
    \bottomrule
  \end{tabular}
\end{table}

%% file: src/7_conclusion.tex
\section{Conclusion}
\label{sec:conclusion}

In this work, we introduced \perfrace{}, a benchmark for long-running performance optimization of real-world software, and \boxbox{}, which combines agent-triggered, full-harness revision from task-specific experience with kernel-managed hot reloading that preserves execution context. On \perfrace{}, offline Harness Evolution yields inconsistent gains, whereas \boxbox{} achieves higher best-so-far correct speedup and greater time and cost efficiency than the evaluated baselines. It achieves a \speedupat{4\mathrm{h}} of \(1.70\times\), compared with \(1.54\times\) for the strongest baseline, and reduces mean model cost by \(29.3\%\) relative to Frozen Harness. These results support harness optimization as an agent capability for improving how it solves a task while continuing its execution.

\section*{Security and Responsible AI Limitations}

\boxbox{} allows an agent to modify executable harness code during an ongoing task, which introduces security and Responsible AI risks beyond those of a fixed harness. In particular, the agent executes and modifies code that may originate from untrusted repositories, and harness revisions may introduce security regressions, weaken existing safeguards, or create unintended execution behavior. More generally, self-modifying agent mechanisms could be misused to alter controls or expand capabilities in ways not intended by system operators. Our evaluation is conducted in isolated research environments with kernel-managed validation and rollback for harness updates and separate shadow verification for task evaluation. These mechanisms reduce experimental risk but do not constitute a complete security boundary or guarantee safe behavior under adversarial conditions. \boxbox{} should therefore be viewed as a research prototype rather than a production-ready self-modifying agent framework. Any production deployment would require additional sandboxing, access control, security review, monitoring, and human oversight of agent-generated harness changes before they are activated.

%% file: src/8a_experiment_appendix.tex
\section{Experimental Details}
\label{app:experiments}

We describe the experimental environment and evaluation details in Appendix~\ref{app:experiments-config}, followed by the experiments for Harness Evolution and Harness Scaling in Appendices~\ref{app:experiments-he} and~\ref{app:experiments-hs}. Appendix~\ref{app:case-study} examines how \boxbox{} harness revisions support task execution, and Appendix~\ref{app:experiments-complete-results} provides additional per-task optimization trajectories.

\subsection{Experimental Setup}
\label{app:experiments-config}

\textbf{Experimental environment.}
To facilitate replication of our experimental environment, we conduct
experiments on an Azure Standard\_D96ads\_v5 virtual machine~\citep{microsoft2026dadsv5}
with 96 vCPUs backed by an AMD EPYC 7763 processor and 384 GiB of system memory,
running Ubuntu 24.04 with Docker.

\textbf{Initial harness.}
All methods start from Terminus-KIRA~\citep{terminuskira2026}, a terminal-based harness that executes
shell commands and returns their outputs to the model. \frozen{} retains
this harness throughout task execution while the agent iteratively optimizes
the project code.

\textbf{Time and cost accounting.}
Figure~\ref{fig:boxbox-main-results} and Table~\ref{tab:boxbox-main-results}
use the same 600 test runs: 50 tasks, four methods, and three runs per
method and task.
For all methods, elapsed task time sums recorded agent-execution and verification phases.
Agent execution includes harness review; environment setup is excluded.
Only evaluations started within the four-hour budget contribute to
\speedupat{4\mathrm{h}}.
We calculate model cost from recorded token usage using the rates in
Table~\ref{tab:model-token-costs}.
Reported mean model costs per test run include all recorded model calls; cost and token
curves report cumulative usage at each task-budget time point.

\begin{table}[H]
  \centering
  \small
  \caption{Model token rates used for cost calculation (USD per million tokens).}
  \label{tab:model-token-costs}
  \begin{tabular}{lrrr}
    \toprule
    Model & Uncached input & Cached input & Output \\
    \midrule
    GPT-5.6-Luna & \$1.00 & \$0.10 & \$6.00 \\
    GPT-5.6-Terra & \$2.50 & \$0.25 & \$15.00 \\
    % Rates from the experiment server's configs/cost-models.toml.
    \bottomrule
  \end{tabular}
\end{table}

\phantomsection
\textbf{Statistical analysis.}
\label{app:experiments-statistics}
The bands in Figure~\ref{fig:boxbox-main-results} show the 2.5th and 97.5th
percentiles of 5,000 resampled benchmark runs over the fixed 50-task set.
Each sample independently selects one of the three observed runs per task
and aggregates the selected scores using the harmonic mean.
The central curves instead use all three runs per task, with the
geometric-then-harmonic aggregation defined in
Equation~\ref{eq:aggregate-speedup}.
In the cost panel, speedup intervals are plotted at the mean cumulative
cost for each task-time point; they do not represent uncertainty in cost.

\subsection{Harness Evolution}
\label{app:experiments-he}

Our \he{} baseline adapts the implementation of \citet{wang2026harnessevolution} to \perfrace{}, retaining its core diagnosis-and-revision workflow. We detail the offline optimization procedure and examine training progress and the generalization of the selected harness to held-out tasks.

\textbf{Setup.} We adapt task execution, evaluation, and experiment orchestration to \perfrace{}, replacing the task-success objective with best-so-far correct speedup under a fixed time budget and tailoring the diagnostic and revision prompts to guide harness improvements that increase speedup and reduce costs.

The reference \he{} experiments use full-batch training on Terminal-Bench 2.1~\citep{wang2026harnessevolution}. Publicly released mini-swe-agent trajectories generated with DeepSeek-V4-Flash on this benchmark~\citep{kwok2026llmasaverifier} average 50 agent steps and 12.2 minutes per run, compared with 965 steps for our \frozen{} runs under a four-hour budget on \perfrace{}. We therefore limit each training run to 45 minutes and partition the 40 training tasks into mini-batches of five to keep the volume of trajectory evidence manageable for diagnosis. Across three reshuffled epochs (24 iterations), we evaluate $h_{i-1}$ on one mini-batch at iteration $i$ and use the resulting trajectories and evaluation outcomes to produce $h_i$. For validation, we compare $h_8$, $h_{16}$, and $h_{24}$ using fresh baseline measurements. Table~\ref{tab:he-configuration-cost} summarizes the \he{} pipeline and its recorded model costs.
% Means over all 445 released TB2.1 runs (89 tasks x 5) and 150 Frozen runs
% (50 tasks x 3). Count agent records only; deduplicate Frozen checkpoint snapshots.
% TB2.1 elapsed time is measured from trial start to finish, not from the budget field.
% TB2.1 source: https://github.com/llm-as-a-verifier/llm-as-a-verifier/tree/628d8fc512c126c4ca8c61dffa44ddba0203c872/data/terminal_bench_2.1_trajs

\begingroup
% Keep spacing around this adjacent table/figure pair fixed under \flushbottom.
\setlength{\intextsep}{6pt}
\begin{table}[H]
  \centering
  \small
  \caption{\he{} configuration and recorded model costs. Training and validation costs include retries and reruns and are lower bounds where usage records are missing.}
  \label{tab:he-configuration-cost}
  \begin{tabular}{@{}lrlrr@{}}
    \toprule
    Stage & Tasks & Evaluated runs & Budget per run & Cost (\$) \\
    \midrule
    Training & 40 & 24 iterations $\times$ 5 tasks & 45 min & 200.34 \\
    Validation & 10 & 3 harnesses $\times$ 3 runs/task & 2 h & 1,176.12 \\
    Test & 50 & h24, 3 runs/task & 4 h & 3,683.34 \\
    \bottomrule
  \end{tabular}
\end{table}

\begin{figure}[H]
  \centering
  % Source: artifacts/plot_he_training_validation.py.
  % Widths follow the panel aspect ratios to keep plot heights aligned.
  \begin{subfigure}[t]{0.6539\textwidth}
    \centering
    \includegraphics[width=\linewidth]{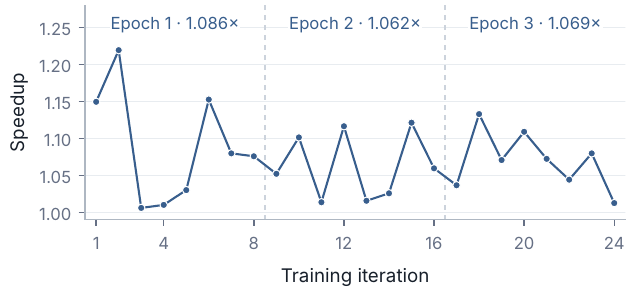}
    \caption{Training progress}
    \label{fig:he-training-progress}
  \end{subfigure}
  \hfill
  \begin{subfigure}[t]{0.3261\textwidth}
    \centering
    \includegraphics[width=\linewidth]{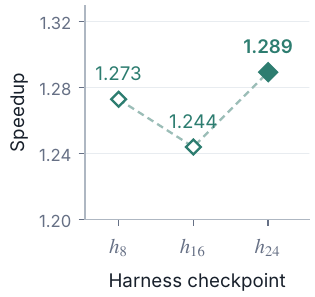}
    \caption{Validation performance}
    \label{fig:he-validation-performance}
  \end{subfigure}
  \caption{\he{} training and validation performance. \subref{fig:he-training-progress} Mini-batch training speedup, with aggregate scores for each epoch. \subref{fig:he-validation-performance} Validation speedup of harness checkpoints $h_8$, $h_{16}$, and $h_{24}$. Speedups are aggregated as in Equation~\ref{eq:aggregate-speedup}.}
  \label{fig:he-training-validation}
\end{figure}
\endgroup

\textbf{Training, validation, and test results.} Training performance does not improve consistently across epochs (Figure~\ref{fig:he-training-progress}). Aggregate speedup decreases from $1.086\times$ in epoch 1 to $1.062\times$ in epoch 2 and recovers to $1.069\times$ in epoch 3, remaining below its initial level. Validation performance does not improve consistently across iterations (Figure~\ref{fig:he-validation-performance}), with $h_{24}$ achieving the highest score among the evaluated checkpoints. We therefore use $h_{24}$ for testing, where it achieves $1.281\times$, below Frozen Harness at $1.442\times$ (Table~\ref{tab:boxbox-main-results}). These results show that additional offline revisions do not reliably improve training performance or yield a test-time advantage, consistent with prior findings of limited transfer~\citep{wang2026harnessevolution}. Including the recorded offline cost raises mean cost to at least \$33.73 per test run.

\textbf{Failure analysis.} To understand \he{}'s failure modes, we examine how a revision introduced during training affected execution on a test task. At iteration 5, diagnostic reports identified two patterns: (1) measurements showed faster execution of the baseline code, which had either remained unchanged or been restored after edits were discarded; and (2) records of faster execution did not specify which code version had been measured. These patterns made it difficult to determine whether measured gains reflected actual code improvements or measurement noise. \he{} therefore added a check to the harness to prevent agents from submitting code without evidence of an actual performance improvement.

Specifically, the agent had to record the current and baseline commits, at least three timing measurements for each version, and correctness evidence in \texttt{.autoresearch/promotion.json}. The harness checked that the record matched the current commit, the worktree was clean, the code differed from the baseline, and the measured improvement exceeded a noise threshold. If these requirements were not met, it blocked submission and returned feedback directing the agent to repair. The check read this record and inspected Git state in the working directory selected by the harness (Figure~\ref{fig:he-failure-analysis}).
% Original diagnostic overview, evolution report, and change manifest:
% exp/artifacts/harness-evolution-h24/iteration-5-diagnosis.md.
% Iteration 5 evaluates h4; its reports are under generations/0004.

\begin{figure}[!htb]
  \begingroup
  \lstset{style=harnessdiff}
  \noindent\textbf{Check the repository and read local evidence.}
  % Selected additions from iteration 5 (e2db056), recurnus/terminus_kira.py.
\begin{lstlisting}[linebackgroundcolor={\DiffBackground{1,3,4,5,6,8,9,10,11,12}{}}]
+ workdir = environment.task_env_config.workdir
  ...
+ head = await environment.exec("git rev-parse HEAD", cwd=workdir, timeout_sec=30)
+ if head.return_code != 0:
+     errors.append(f"cannot resolve candidate HEAD: {head.stderr.strip()}")
+     return errors
  ...
+ record = await environment.exec(
+     "cat -- .autoresearch/promotion.json",
+     cwd=workdir,
+     timeout_sec=30,
+ )
\end{lstlisting}
  \noindent\textbf{Check the measured improvement.}
  % Selected additions from e2db056, recurnus/materiality.py, validate_promotion_record.
\begin{lstlisting}[linebackgroundcolor={\DiffBackground{1,2,4,5}{}}]
+ threshold = max(0.05 * baseline_median, 2 * baseline_noise)
+ improvement = baseline_median - candidate_median
  ...
+ if improvement < threshold:
+     errors.append(f"{PROMOTION_MARGIN_ERROR_PREFIX}: need at least {threshold} seconds")
\end{lstlisting}
  \noindent\textbf{Block submission when checks fail.}
  % Selected additions from e2db056, recurnus/terminus_kira.py.
\begin{lstlisting}[linebackgroundcolor={\DiffBackground{1,2,3,4,5,6,7,8,9,10}{}}]
+ if phase_outcome.action is not PhaseAction.CONTINUE:
+     gate_errors = await self._promotion_gate_errors(self._session)
+     if gate_errors:
+         phase_outcome = PhaseOutcome()
+         promotion_gate_feedback = (
+             "WARNINGS: PROMOTION GATE BLOCKED; no verification was requested.\n"
+             + "\n".join(f"- {error}" for error in gate_errors)
+             + "\nCreate or repair .autoresearch/promotion.json, then retry completion."
+         )
+         feedback = "\n".join(filter(None, (feedback, promotion_gate_feedback)))
\end{lstlisting}
  \endgroup
  \caption{Selected additions to \he{}'s harness at iteration 5. The harness checks the repository and performance evidence before submission, blocking submission and returning repair instructions when a check fails.}
  \label{fig:he-failure-analysis}
\end{figure}

The check enforced fixed requirements based on failures observed during training. However, its assumptions about repository location did not hold for some test tasks, causing it to block submissions before assessing the performance evidence. To examine this failure pattern, we selected \texttt{mirage--kubernetes-triage} for detailed trajectory analysis.

In the selected run, the agent prepared Git commits and performance evidence inside the editable repository, while the harness ran its submission checks in the parent workspace. The check therefore failed to resolve the candidate commit and returned \texttt{cannot resolve candidate HEAD}, blocking submission before assessing the performance evidence. Repeated attempts to repair the evidence in the repository did not change the directory used by the harness. The trajectory records 473 blocked submission requests spanning approximately 3 hours and 46 minutes.

This case illustrates both challenges in Section~\ref{sec:introduction}. Difficulty in attributing measured gains motivated stricter submission requirements, but did not establish that these requirements would help the agent find better optimizations. On the selected test task, the check's repository-location assumptions instead led to repeated repair attempts and blocked submissions, illustrating how a revision intended to address training failures can hinder execution under new task-specific conditions.
% Evidence: mirage--kubernetes-triage__my56nR7, h24 test-0, final-delivery
% trajectory.json. Observation steps matching both "PROMOTION GATE BLOCKED"
% and "cannot resolve candidate HEAD": 473, from step 43 to step 1005.
% First/last timestamps: 2026-09-24T15:53:15.273016+00:00 and
% 2026-09-24T19:38:49.848844+00:00 (wall-clock span, not official budget time).
% Step 53 checks local packet identity and reports a clean Git status;
% step 54 is blocked again. This local check is not an official Judge verdict.

\subsection{Harness Scaling}
\label{app:experiments-hs}

Our \hs{} baseline adapts the implementation of \citet{wang2026harnessevolution} to \perfrace{}. We describe the execution procedure, analyze model costs, and examine a failure case.

\textbf{Setup.} We adapt task execution and evaluation to \perfrace{}, using best-so-far correct speedup under a fixed time budget as the optimization objective. Each run shares a four-hour budget across task execution, harness optimization, and evaluation. An external controller revises the harness using feedback from each completed attempt and starts a new agent session, preserving project-code progress and resetting the execution context.

We record 2,132 \hs{} attempts, with a median of 12 per run. Across 1,983 boundaries, the median interval between one attempt ending and the next beginning is 3.37 minutes, including harness optimization and runtime transition.
% Evidence: canonical HS jobs under /mnt/perfbench/jobs/harness-scaling/
% on azureuser@20.3.99.110, waves:
% harness-scaling-test0-wave-0-restart-r2-c38bc18b,
% harness-scaling-test0-wave-1-restart-r3-c38bc18b,
% harness-scaling-test0-wave-2-restart-r1-6f0724fc,
% harness-scaling-test0-wave-3-restart-r1-8574a5cb.
% agent/harness-scaling/campaign-summary.json: 149 summaries, 2132 rounds;
% kitops--minilm-pack__zPeYG38 has no summary. Median rounds/run = 12.
% Across 1983 adjacent rounds, median (next.started_at - current.completed_at)
% = 202.418758869 seconds. Evolution begins after attempt completion.

\textbf{Model costs.}
\hs{}'s session resets can lower input-token costs by reducing the accumulated conversation history. Although project-code progress is retained, subsequent attempts may need to reconstruct earlier decisions or repeat experiments whose outcomes are no longer available in the execution context. \boxbox{} preserves this context across harness updates, allowing subsequent optimization to build directly on earlier findings. This trade-off helps explain why \hs{}'s lower total model cost does not necessarily translate into greater optimization progress per dollar.

\begin{figure}[!htb]
  \centering
  % Source: artifacts/plot_hs_case_study.py.
  \includegraphics[width=\linewidth]{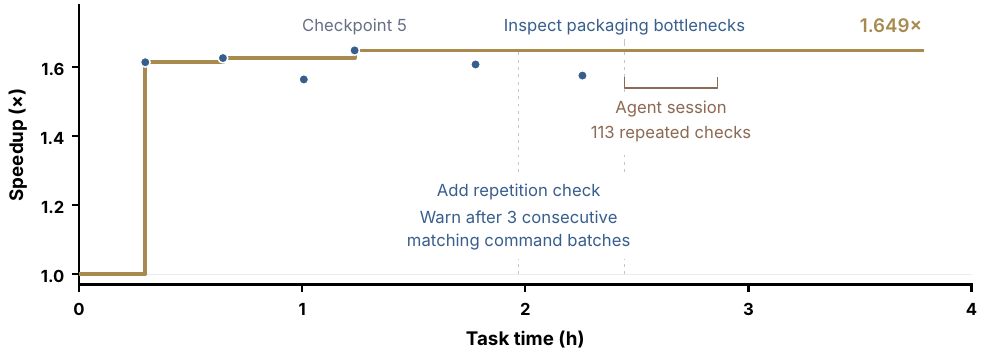}
  \caption{\hs{} optimization trajectory on \texttt{kitops--minilm-pack}. Annotations show selected harness revisions and an agent session containing 113 consecutive checks that produced no new performance measurements.}
  \label{fig:hs-kitops-trajectory}
\end{figure}

\textbf{Failure analysis.} To examine \hs{}'s optimization process, we trace one \texttt{kitops--minilm-pack} run (Figure~\ref{fig:hs-kitops-trajectory}). During one harness optimization step, \hs{}'s controller reported that the agent repeatedly rebuilt the project, ran tests, and reran the workload on the same code version. It therefore added a repetition check to warn after three consecutive matching command batches.

However, the new repetition check did not stop the agent from performing the same checks using different command forms. The agent ran 113 consecutive commands repeating the same checks, without modifying code or collecting new performance measurements. The check only looked for consecutive repetitions of the same commands, so it could miss repeated checks written in different ways. The run reached its best correct speedup of $1.649\times$ at checkpoint 5, with no further improvement in subsequent recorded checkpoints. The repeated checks revealed a need for further harness revision during execution, yet \hs{} performs harness optimization only between agent sessions, leaving the current session to continue under the existing harness.

This case illustrates how session-level external control can delay harness revisions needed during ongoing task execution. The revised check missed repeated validation expressed through different commands, and the agent continued spending task time without making code changes or collecting new performance measurements. Allowing the agent to initiate further harness revision when these limitations become apparent would provide an earlier opportunity to address them.
% Case evidence: canonical wave 0 / kitops--minilm-pack__zPeYG38.
% agent/harness-scaling/rounds/0006 and 0009:
% evolve/change_manifest.json, target/terminal.txt, target/context.json.
% Round 6: c2fe4347 -> f2a361a2 adds the consecutive-repeat warning.
% Round 9 runs fe78b560 with that check retained; e9c0fbc3 adds recent-history blocking.
% The round-9 terminal markers 92--204 identify 113 consecutive commands
% reading the worker setting and checking the executable; all exit with status 0.
% Both attempt contexts record attempt_timeout_sec=1500 and attempt_timed_out=true.
% Figure 1 archive: harness-scaling-test0.json.gz, baseline=1.238681459,
% checkpoint-05=0.751173639; no subsequent accepted checkpoint improves on it.
% Result SHA256: 58d7447bb0d907291af98d6e2f26eab78f263e13073fd94d2ca2e55590e3a0ce.
% Round 10 lacks terminal evidence because the container was paused;
% its timeout does not establish whether the new blocking check was effective.

\FloatBarrier
% Evidence: exp/recurnus-historical-edits/case-studies.md.
\subsection{Case Study}
\label{app:case-study}
\begingroup
\lstset{style=harnessdiff}

To analyze how \boxbox{} supports effective task execution, we examine two complementary aspects of harness optimization. Appendix~\ref{app:case-trajectories} traces how agent-triggered reviews address difficulties encountered during ongoing tasks. Appendix~\ref{app:case-modifications} examines how full-harness editing can support task performance and reduce model costs through changes to agent guidance, tools, and execution behavior.

\subsubsection{Harness Optimization Trajectories}
\label{app:case-trajectories}

We examine two four-hour runs, tracing the difficulties encountered during task execution, the resulting harness revisions, and their use in subsequent optimization.

\textbf{CCC.} To reduce CCC's compilation time on SQLite, the agent profiles compiler stages and tests code changes through repeated compilation runs (Figure~\ref{fig:boxbox-ccc-trajectory}). During the first review, the agent identified repeated manual processing of timing output and added a helper for repeated runs, result recording, and candidate comparisons. Using this helper, the agent established a baseline and evaluated compiler changes, but still had to extract and compare individual stage timings manually. To automate this step, the agent added stage-timing summaries to the helper in the second review, then used them to investigate optimization passes and code generation.

Later, a change to CCC's source code appeared to reduce compilation time by 9.4\% in separately measured batches. Yet when the agent alternated runs of the modified and unchanged compiler versions, their median compilation times were nearly identical, suggesting that the apparent gain came from measurement variation between batches rather than the code change. The agent therefore incorporated this comparison procedure into the helper during the third review to improve measurement reliability, then used it to reject several compiler changes that showed no consistent speedup. Continuing with this helper, the agent developed mnemonic-parsing and operand-splitting changes that reduced official running time from $6.321$ to $6.095$ and then $5.792$ seconds.

\begin{figure}[!htb]
  \centering
  % Source: artifacts/plot_boxbox_ccc_case_study.py.
  \includegraphics[width=\linewidth]{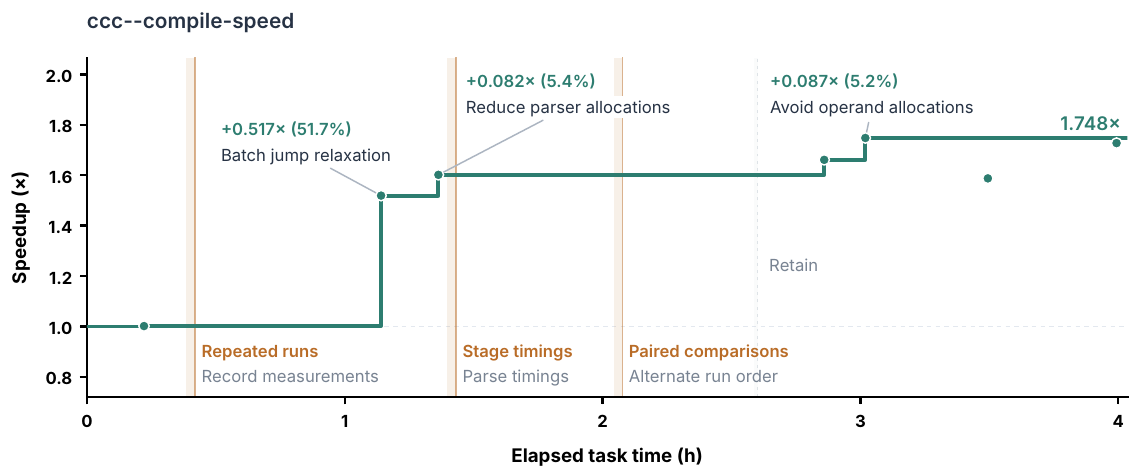}
  \caption{\boxbox{} optimization trajectory on \texttt{ccc--compile-speed}. Points show correct evaluation results and the line shows best-so-far speedup. Annotations mark harness review outcomes and task-code changes; numerical gains are absolute and relative increases over the preceding best-so-far speedup.}
  \label{fig:boxbox-ccc-trajectory}
\end{figure}

% CCC source: recurnus-evidence-scope-test25-3x4h-c25-20260912 /
% ccc--compile-speed__az4MaTZ. Task trajectory sequences 70/98 show initial
% helper use; 274--275 show phase
% summaries and their use; 360--363 show the batch/alternating discrepancy;
% 370--371 show paired-ledger execution; 452/467 commit later improvements;
% checkpoint-06 and sequence 513 record the local/official disagreement.
% Review 0004, step 10, records retention after the paired helper was used
% to reject several candidates without consistent gains.

\textbf{FluidSynth.} To optimize FluidSynth's MIDI-to-WAV rendering performance, the agent tests code changes through repeated workload execution and output checks. Figure~\ref{fig:boxbox-fluidsynth-trajectory} traces harness revisions during this process. The first review noted that long-running commands required extra model calls to continue waiting, but retained the harness because a proposed waiting tool would duplicate existing functionality. As this pattern recurred, the second review extended waiting inside the existing command runner. During subsequent audio-write batching experiments, a single tool call completed a build, five renders, and output checks, allowing the agent to assess the candidate without an additional waiting call.

Comparing batch sizes still required repeated setup of builds, timing loops, and output checks. The third review therefore added \texttt{benchmark\_matrix}, reusing the updated command runner to execute labeled experiments and report timings and exit status. The agent then used it for six repeated renders with output-hash checks and subsequent configuration comparisons. Rendering performance improved further after the agent disabled OpenMP in the build configuration. The fourth review retained the harness, judging its existing capabilities sufficient after the agent resolved a build-state issue through a clean rebuild. The run reached a best-so-far speedup of $1.183\times$.

\begin{figure}[!htb]
  \centering
  % Source: artifacts/plot_boxbox_case_study.py.
  \includegraphics[width=\linewidth]{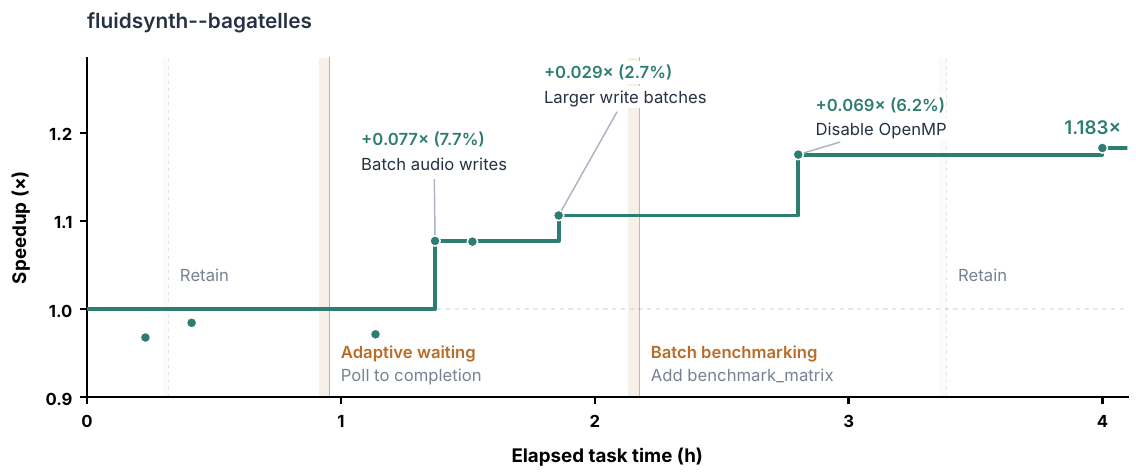}
  \caption{\boxbox{} optimization trajectory on \texttt{fluidsynth--bagatelles}, using the plotting conventions of Figure~\ref{fig:boxbox-ccc-trajectory}. Reviews extend command waiting and add batch benchmarking.}
  \label{fig:boxbox-fluidsynth-trajectory}
\end{figure}

Together, these cases illustrate how harness optimization can improve task performance and reduce model costs by improving the agent's performance optimization workflow, even though the reviews do not directly modify project code.

% Keep trajectory figures with their analysis while allowing text to fill the page.
\FloatBarrier
\subsubsection{Representative Harness Modifications}
\label{app:case-modifications}
\newcounter{harnesscase}

We examine five representative harness updates that change agent guidance, tools, or execution behavior. While Figure~\ref{fig:boxbox-review-analysis-modifications} reports the prevalence of modification categories, Table~\ref{tab:harness-case-overview} summarizes the concrete changes analyzed below.

\begin{table}[!htb]
  \centering
  \caption{Representative harness updates grouped by modification type. Case numbers refer to the detailed analyses below; task IDs follow the catalog in Table~\ref{tab:perfrace-task-catalog}.}
  \label{tab:harness-case-overview}
  \small
  \setlength{\tabcolsep}{4pt}
  \renewcommand{\arraystretch}{1.2}
  \begin{tabularx}{\textwidth}{@{}cl>{\raggedright\arraybackslash}p{0.29\textwidth}>{\raggedright\arraybackslash}X@{}}
    \toprule
    Case & Category & Task ID & Modification \\
    \midrule
    \ref{case:feedback} & Review feedback & {\scriptsize\path{n8n-mcp--tool-batch-speed}} &
      Redirect the next experiment toward SQLJS conversion and statement overhead. \\
    \ref{case:prompt} & \shortstack[l]{Prompt rules\\and skills} & {\scriptsize\path{sum--text-attest}} &
      Require alternating A/B measurements when apparent gains are close to noise. \\
    \ref{case:polling} & Tool functionality & {\scriptsize\path{rust-clippy--cargo-package-lint}} &
      Extend internal command polling to reduce model-directed waiting turns. \\
    \ref{case:context} & Execution logic & {\scriptsize\path{trixi-framework-trixi-jl--self-gravitating-sedov-amr}} &
      Add a message-count-based context-compaction path. \\
    \ref{case:fault} & Fault fix & {\scriptsize\path{rust-clippy--cargo-package-lint}} &
      Replace an exception on shell exit with a session-recovery call. \\
    \bottomrule
  \end{tabularx}
\end{table}

The following cases proceed from changes in agent guidance to changes in executable behavior. Cases~\ref{case:feedback}--\ref{case:prompt} examine review feedback and prompt-based guidance for experiment selection and evaluation. Case~\ref{case:polling} examines internal command polling to reduce model-directed waiting. Cases~\ref{case:context}--\ref{case:fault} turn to context management and terminal recovery within the agent loop. Together, they illustrate how full-harness editing can affect both the direction of the search and the process used to carry it out.

% n8n-mcp--tool-batch-speed__HF3CmJx:transition:0001
\par\medskip
\refstepcounter{harnesscase}
\label{case:feedback}
\noindent\textbf{Case~\theharnesscase. Review feedback: redirecting an experiment}
\par\nopagebreak

\noindent\textbf{Task:} {\small\path{n8n-mcp--tool-batch-speed}}

\textbf{Background.}
The task is to reduce the end-to-end running time of the n8n-mcp server on a fixed batch of workflow-building tool calls while preserving response correctness. The agent attempted to reduce logging overhead, but local measurements showed no improvement. The review therefore used profiling evidence to identify a different direction for the next experiment.

\noindent\begin{minipage}{\linewidth}
\begin{lstlisting}[linebackgroundcolor={\DiffBackground{1,2,3,4,5,6,7,8}{}}]
+ The remaining profiler clue is the SQLJS adapter's repeated `all/get`
+ row/object conversions and statement lifecycle overhead. Treat this as
+ a hypothesis, not a conclusion: inspect the exact hot path and design
+ one general correctness-preserving experiment ...
+
+ ... Use a separate branch from main, build, run the exact workload with
+ paired repeated clean-vs-experiment medians, and compare response hashes.
+ Reject changes within noise or with any response mismatch.
\end{lstlisting}
\end{minipage}

\textbf{Review feedback.}
The feedback directs the agent to stop revisiting logging and investigate SQLJS data-conversion and statement-management overhead. It also specifies an experimental workflow, asking the agent to preserve the current project-code version and test modifications on a separate branch through paired repeated workload measurements and response-hash comparisons. Modifications should be rejected if their measured gains fall within measurement noise or their responses differ.

\textbf{Discussion.}
This case illustrates how review feedback can support performance optimization by guiding both experiment selection and evaluation. Redirecting effort away from an unsuccessful direction can reduce repeated exploration, while paired measurements and response checks help the agent distinguish useful code changes from measurement noise or incorrect behavior.

\par\medskip
\refstepcounter{harnesscase}
\label{case:prompt}
\noindent\textbf{Case~\theharnesscase. Prompt rules and skills: comparing noisy measurements}
\par\nopagebreak

\noindent\textbf{Task:} {\small\path{sum--text-attest}}
% sum--text-attest__vx33d4E:transition:0008

\textbf{Background.}
The task is to reduce the running time of SUM's batch text-attestation workflow while preserving its outputs. Across successive experiments, the agent encountered apparent gains that did not hold under more controlled measurements. Earlier reviews had already warned against drawing conclusions from warmup effects and uncontrolled measurement order, yet a later code change again appeared promising based on a single fast run. To address this recurring problem, the review incorporated the measurement guidance into the main prompt for use across subsequent experiments.

\begin{lstlisting}[linebackgroundcolor={\DiffBackground{1,2,3,4,5,6,7}{}}]
+ When a candidate's apparent gain is comparable to run-to-run noise, do not
+ infer from one fast first run. Keep the unchanged `main` reference and
+ candidate branch, warm both in the same controlled procedure, then run
+ alternating fresh-process A/B repetitions with identical output checks
+ and compare medians (and spread) before accepting or rejecting the hypothesis.
+ Align this local procedure with the official workload boundary before
+ using a tiny timing difference to justify submission.
\end{lstlisting}

\textbf{Harness update.}
The prompt rule instructs the agent to keep an unchanged reference version and test modified code on a separate branch. It requires the same warmup procedure for both versions, followed by alternating fresh-process workload runs with identical output checks. The agent should compare median running times and their spread before accepting or rejecting the modification, with the measurement procedure aligned with the official workload.

\textbf{Discussion.}
This case illustrates how prompt revisions can provide consistent guidance for evaluating code changes across successive experiments. More reliable comparisons can help the agent distinguish genuine improvements from measurement noise, reducing the risk of retaining ineffective changes or discarding useful ones based on unstable timing evidence.

\par\medskip
\refstepcounter{harnesscase}
\label{case:polling}
\noindent\textbf{Case~\theharnesscase. Tool functionality: reducing model calls for command polling}
\par\nopagebreak

\noindent\textbf{Task:} {\small\path{rust-clippy--cargo-package-lint}}
% rust-clippy--cargo-package-lint__cuUYeRH:transition:0001

\textbf{Background.}
During performance optimization on Rust Clippy, the agent builds the project and runs benchmarks to evaluate code changes. The command tool could return before these operations finished, requiring additional model calls simply to continue waiting. The review therefore extended the tool's internal waiting behavior to reduce these repeated interactions.

\textbf{Harness update.}
The update extends internal waiting to 180 seconds for ordinary commands, subject to the existing block timeout. The tool continues checking for the completion marker and returns as soon as it appears, or provides a progress snapshot when the waiting window expires.

\begin{lstlisting}[linebackgroundcolor={\DiffBackground{1,3,4,5,6,9}{8}}]
+ AUTO_COMMAND_WAIT_SEC = 180
  ...
+ wait_budget = min(
+     BLOCK_TIMEOUT_SEC,
+     max(command.duration_sec, AUTO_COMMAND_WAIT_SEC),
+ )
  ...
- while time.monotonic() - start < command.duration_sec:
+ while time.monotonic() - start < wait_budget:
\end{lstlisting}

The tool description is updated alongside the implementation, instructing the agent to request an additional wait only when the command outlasts the new window.

\begin{lstlisting}[linebackgroundcolor={\DiffBackground{3,4,5,6,7}{1,2}}]
- "It is always possible to wait again if the prior output has not finished, "
- "by running empty keystrokes with a duration on subsequent requests to wait longer. "
+ "The harness automatically continues waiting when the completion marker has "
+ "not appeared, up to 180 seconds per tool call, and returns a progress snapshot "
+ "at that bound. "
+ "Use an empty-keystroke wait only when a command exceeds that bound; "
+ "use C-c to interrupt foreground work. "
\end{lstlisting}

\textbf{Discussion.}
This case illustrates how full-harness editing allows agents to adjust tool behavior adaptively to reduce model calls spent solely on waiting. When an unfinished command returns control to the agent, another model call processes the conversation context and generates a continuation request, incurring input- and output-token costs. Internal polling checks command status without invoking the model, allowing commands that finish within the extended window to return their results without these additional calls. Fewer waiting turns also add fewer messages to subsequent context. The potential cost reduction therefore comes from reducing model interaction during command execution, while commands exceeding the window may still require further calls.

\par\medskip
\refstepcounter{harnesscase}
\label{case:context}
\noindent\textbf{Case~\theharnesscase. Execution logic: context compaction}
\par\nopagebreak

\noindent\textbf{Task:} {\small\path{trixi-framework-trixi-jl--self-gravitating-sedov-amr}}
% trixi-framework-trixi-jl--self-g__ZV6MxPM:transition:0003

\textbf{Background.}
During performance optimization on Trixi, the agent accumulated a long interaction history through profiling and repeated experiments. The review reported that late model requests contained approximately 145,000 input tokens, making repeated processing of the accumulated history a substantial source of model cost. The existing summarization mechanism had not sufficiently limited this growth, motivating a revision to the agent loop's context-management logic.

\begin{lstlisting}[linebackgroundcolor={\DiffBackground{1,2,3,4,5,6,7,8,9,10,11,12,13}{}}]
+ if (
+     original_instruction
+     and len(chat.messages) >= 64
+     and len(chat.messages) >= self._last_forced_compaction_messages + 24
+ ):
+     self._last_forced_compaction_messages = len(chat.messages)
+     proactive_summary_result = await self._force_proactive_compaction(
+         chat, original_instruction, self._session,
+     )
+     if proactive_summary_result:
+         prompt, subagent_refs = proactive_summary_result
+         self._pending_subagent_refs = subagent_refs
+         self._pending_handoff_prompt = prompt
  if original_instruction and self._enable_summarize:
      ...
\end{lstlisting}

\textbf{Harness update.}
The update adds a message-count trigger before the existing optional summarization check. When the original task instruction is available, the loop attempts compaction if the history contains at least 64 messages and its message count is at least 24 higher than the count recorded at the last attempt. It records the current count before invoking the existing summarizer through a helper with a timeout. Successful results supply the continuation prompt and subagent references; if summarization fails, execution continues with the existing history.

\textbf{Discussion.}
This case illustrates how full-harness editing allows agents to adapt context-management logic during ongoing tasks. By summarizing accumulated interaction history, the revised loop can reduce the amount of context processed in subsequent model calls and the associated input-token costs.

\par\medskip
\refstepcounter{harnesscase}
\label{case:fault}
\noindent\textbf{Case~\theharnesscase. Fault fix: terminal recovery}
\par\nopagebreak

\noindent\textbf{Task:} {\small\path{rust-clippy--cargo-package-lint}}
% rust-clippy--cargo-package-lint__FnUrnTW:transition:0004

\textbf{Background.}
While the agent was waiting for a long Rust Clippy build, the persistent terminal shell exited. At the next agent-loop entry, the harness raised an exception before its existing session-recovery helper could run. This failure prompted a review of how the harness handles terminal loss during ongoing execution.

\begin{lstlisting}[linebackgroundcolor={\DiffBackground{3,4,5,6}{1,2}}]
- if not await self._with_block_timeout(self._session.is_session_alive()):
-     raise RuntimeError("The persistent terminal shell exited")
+ # The previous episode may have outlived the persistent shell.
+ # Restart it for the next model turn without replaying any
+ # command whose completion was uncertain.
+ await self._with_block_timeout(self.ensure_agent_session())
\end{lstlisting}

\textbf{Harness update.}
The update replaces the exception with a call to the existing session-recovery helper, which preserves a live terminal or starts a replacement if the shell has exited. This prepares a usable terminal for the next model turn without automatically replaying commands whose completion is uncertain. Such commands may already have modified project files, so replaying them could duplicate their effects.

\textbf{Discussion.}
This case illustrates how full-harness editing allows agents to revise error-handling logic when runtime faults interrupt ongoing tasks. By restoring terminal access, the revised harness can enable the agent to inspect the current task state and resume performance optimization after a shell failure.

Together, these cases illustrate how full-harness editing lets agents adapt guidance, tools, and execution logic to the needs of ongoing tasks. Changes to experiment selection and measurement can support task performance by focusing effort on promising hypotheses and improving the reliability of candidate evaluation. Changes to tool execution and context management can reduce model costs by avoiding unnecessary waiting calls and repeated transmission of long histories, while fault-handling revisions support continued execution.
\endgroup

\FloatBarrier
\subsection{Per-Task Optimization Trajectories}
\label{app:experiments-complete-results}

\begin{figure}[p]
  \centering
  \includegraphics[width=\textwidth,height=0.92\textheight,keepaspectratio]{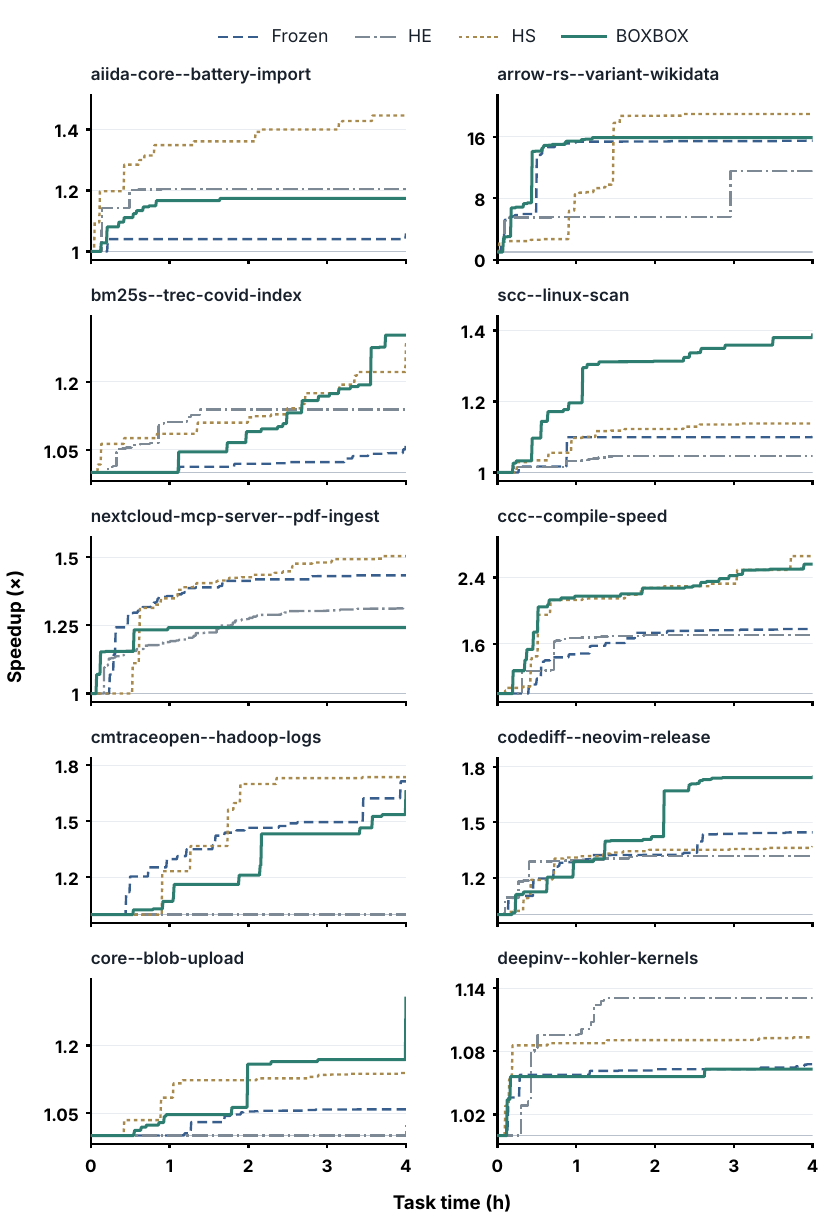}
  \caption{Per-task optimization trajectories (1/5).}
  \label{fig:boxbox-per-task-1}
\end{figure}

\begin{figure}[p]
  \centering
  \includegraphics[width=\textwidth,height=0.92\textheight,keepaspectratio]{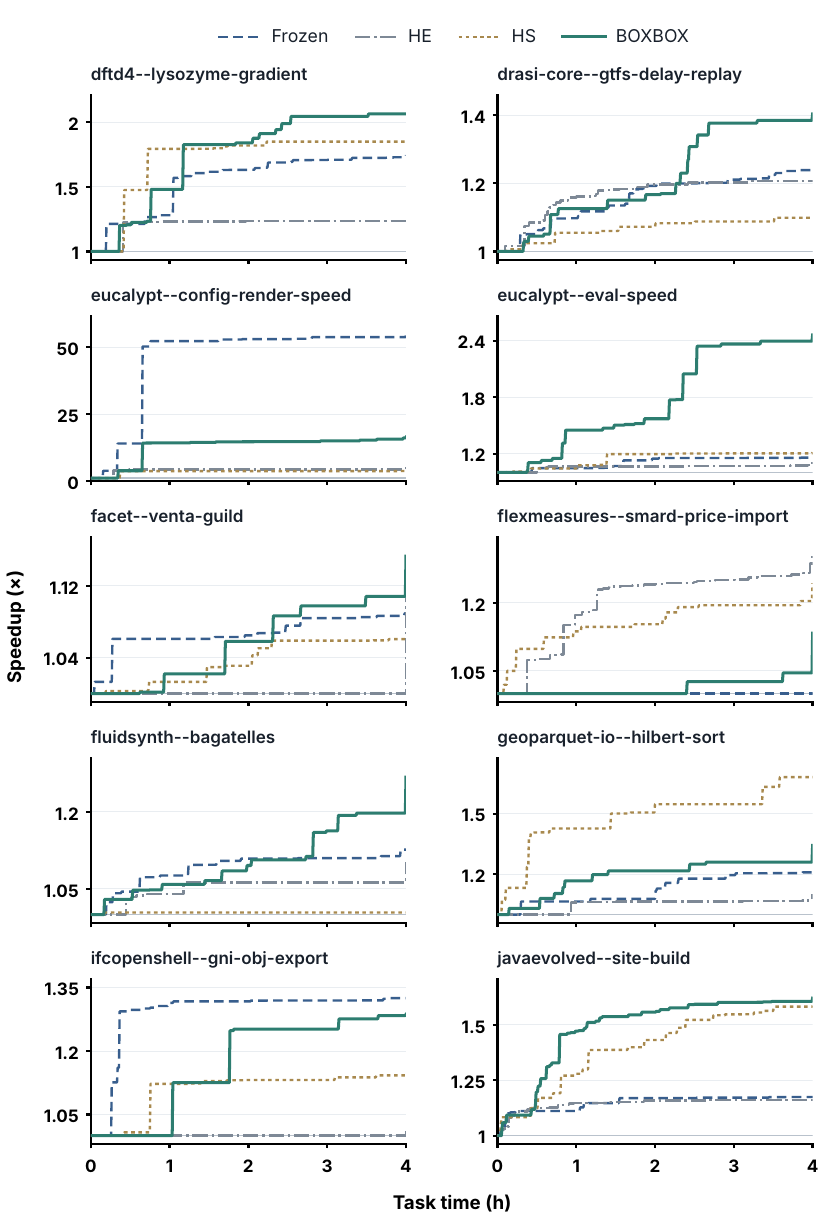}
  \caption{Per-task optimization trajectories (2/5).}
  \label{fig:boxbox-per-task-2}
\end{figure}

\begin{figure}[p]
  \centering
  \includegraphics[width=\textwidth,height=0.92\textheight,keepaspectratio]{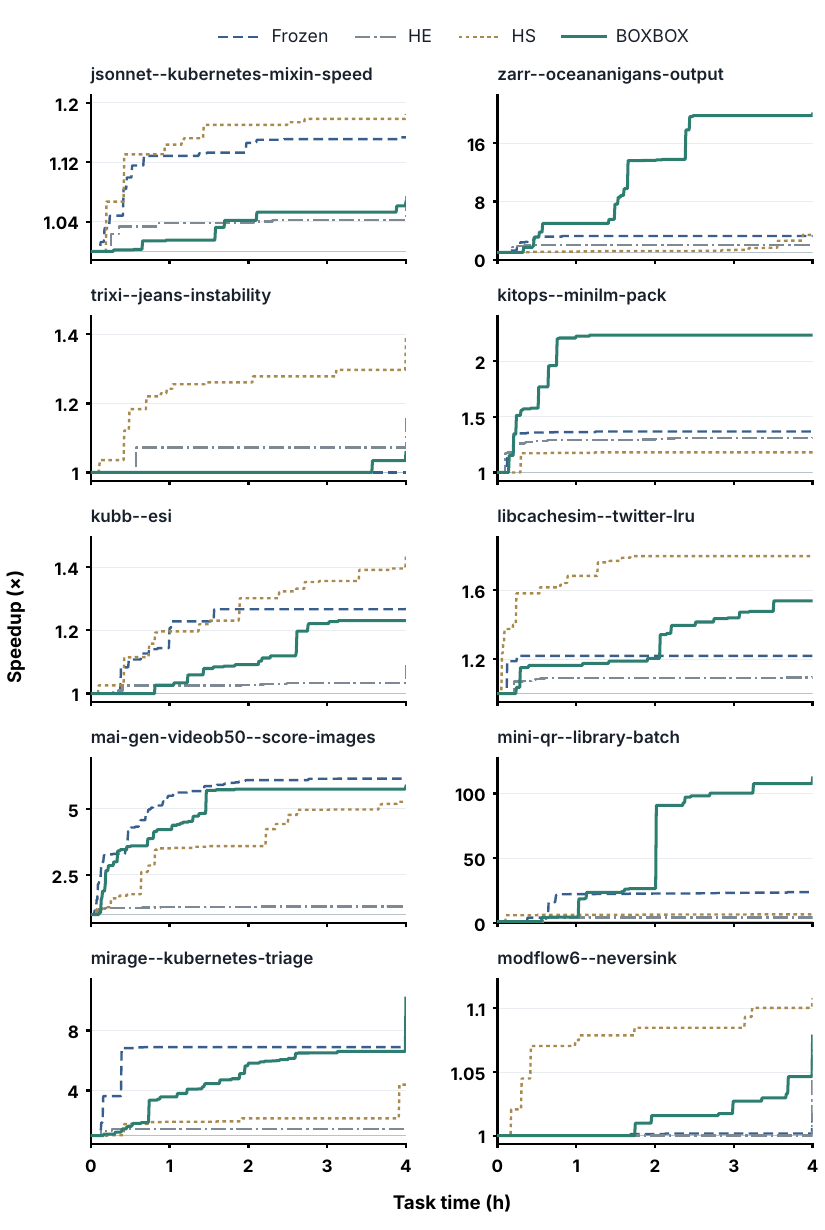}
  \caption{Per-task optimization trajectories (3/5).}
  \label{fig:boxbox-per-task-3}
\end{figure}

\begin{figure}[p]
  \centering
  \includegraphics[width=\textwidth,height=0.92\textheight,keepaspectratio]{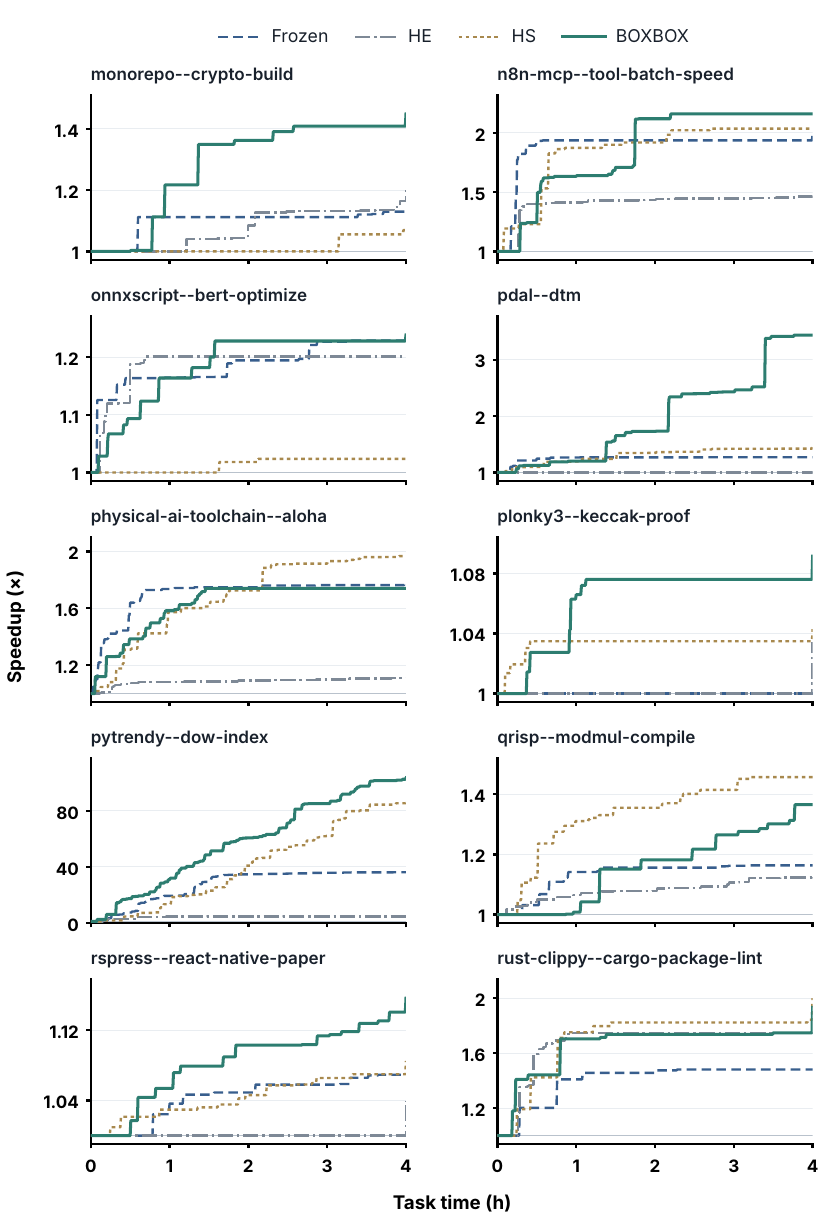}
  \caption{Per-task optimization trajectories (4/5).}
  \label{fig:boxbox-per-task-4}
\end{figure}

\begin{figure}[p]
  \centering
  \includegraphics[width=\textwidth,height=0.92\textheight,keepaspectratio]{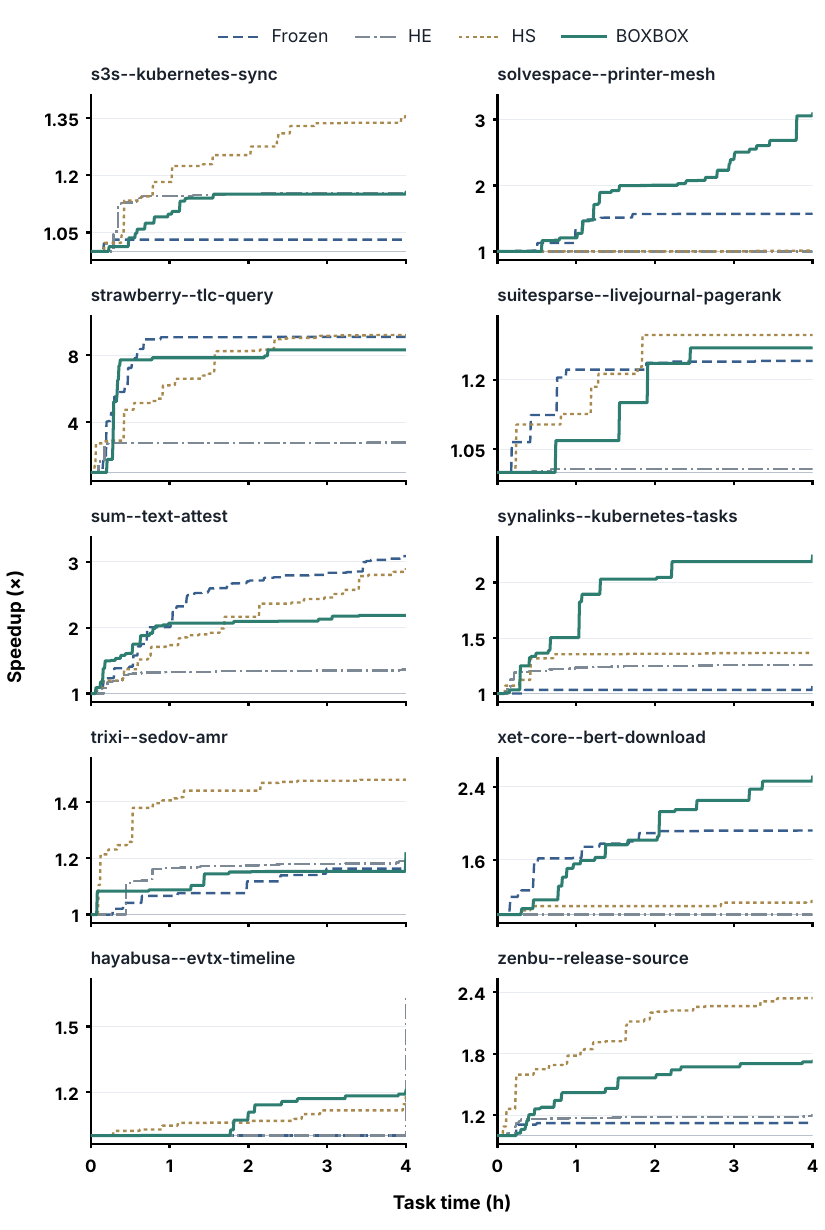}
  \caption{Per-task optimization trajectories (5/5).}
  \label{fig:boxbox-per-task-5}
\end{figure}

Figures~\ref{fig:boxbox-per-task-1}--\ref{fig:boxbox-per-task-5}
show optimization trajectories for all 50 test tasks.
Each curve geometrically averages three runs, following the scoring and
time-accounting conventions of Figure~\ref{fig:boxbox-main-results}.
The vertical scale varies by task.

\clearpage

\clearpage

\clearpage

\clearpage

\clearpage

%% file: src/8b_perfrace_appendix.tex
\section{\perfrace{}: Benchmark Details}
\label{app:perfrace}

The \perfrace{} appendix is organized as follows: Section~\ref{app:perfrace-comparison} compares \perfrace{} with related benchmarks; Section~\ref{app:perfrace-characterization} characterizes its 100 tasks; Section~\ref{app:perfrace-curation-statistics} reports construction and curation statistics; Section~\ref{app:perfrace-pair-verification} describes the Pair Verification protocol and quality control; and Section~\ref{app:perfrace-evaluation} details the evaluation protocol and analyzes gains from longer optimization budgets.

\subsection{Benchmark Curation and Evaluation}
\label{app:perfrace-comparison}

Table~\ref{tab:perfrace-benchmark-comparison} compares \perfrace{} with
representative ultra-long-horizon agent benchmarks and code-optimization
benchmarks along task type, task construction, and evaluation procedure.
Existing ultra-long-horizon benchmarks largely assemble fixed suites of
hand-crafted or expert-curated tasks. In contrast, \perfrace{} combines live
GitHub discovery with agent--human Pair Verification to continually construct
validated optimization tasks from real-world projects and workloads.
Compared with existing code-optimization benchmarks, it does not derive tasks
from known performance patches and evaluates progress repeatedly rather than
only at task completion.

A workload specifies the inputs and operations used to exercise a project.
A microbenchmark measures a narrowly scoped part of that work, such as a
selected API call or operation. For example, SWE-fficiency includes a
performance workload that repeatedly times a NumPy \texttt{einsum}
call~\citep{ma2025swefficiency}.
An end-to-end workload instead measures a complete project-use workflow.
\perfrace{} adopts this broader measurement boundary, leaving agents to
profile the workflow and identify which components to optimize.
The distinction is therefore the scope of the measured computation:
targeted operations versus an end-to-end workflow, rather than whether
the benchmark supplies a workload.

\input{tables/perfrace-comparison}

\subsection{Detailed Benchmark Composition}
\label{app:perfrace-characterization}

We characterize the 100 \perfrace{} tasks by project purpose, primary programming language, license, packaged source size, file count, and fixed workload.

\textbf{Projects.}
The projects represented in \perfrace{} span 16 primary programming languages:
Rust (25 tasks), Python (18), TypeScript (15), C++ (11), C (6), C\# (6), Julia
(5), Go (4), Java (3), Scala (1), Lua (1), OpenSCAD (1), Solidity (1), MATLAB
(1), JavaScript (1), and Zig (1).
The packaged project sources range from 42 to 15,869 files and from 471.49 KiB
to 689.08 MiB, with medians of 886 files and 14.56 MiB.

\begin{figure}[H]
  \centering
  \includegraphics[width=\textwidth]{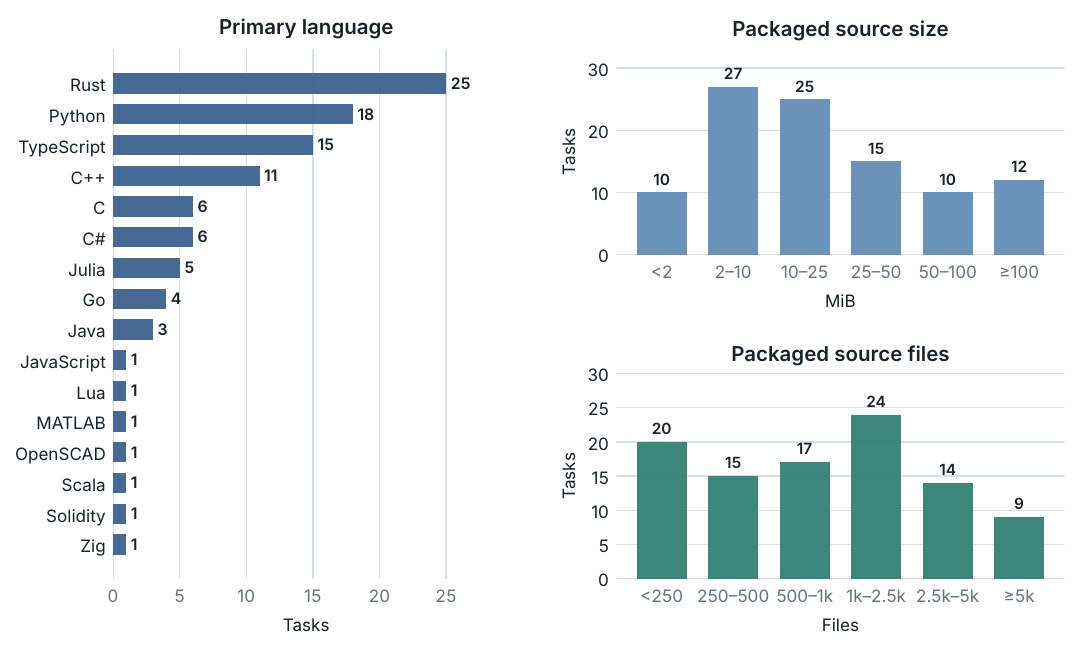}
  \caption{Distribution of primary programming languages, packaged source sizes, and packaged source file counts across \perfrace{} projects.}
  \label{fig:perfrace-projects}
\end{figure}

\textbf{Task catalog.}
Table~\ref{tab:perfrace-task-catalog} lists each task's project, purpose,
license, and fixed workload.

\input{tables/perfrace-task-catalog}

\subsection{Construction and Curation Statistics}
\label{app:perfrace-curation-statistics}

This section describes the implementation of Discovery and Filter and reports the candidate counts and curation effort at each construction stage.

\textbf{Discovery implementation.}
We query the ClickHouse \texttt{github\_events} table for opened pull requests and branch pushes by a configured list of coding-agent accounts, then group matching events by repository.
For each repository, we retain the latest matching pull-request or branch URL and use the GitHub GraphQL API to record its canonical identity, default-branch revision, stars, primary language, license, disk usage, and repository status.

The resulting records are deduplicated by GitHub repository, and projects are excluded when their repository, default branch, or recorded source revision cannot be retrieved.
Each step writes its results incrementally, allowing interrupted collection and metadata retrieval to resume without repeating completed work.

From GitHub activity recorded between May and August 2026, we identified 118,517 repositories, retrieved metadata for 104,428, and retained 103,475 project candidates after these checks.

\input{tables/perfrace-discovery-statistics}

\textbf{Filter implementation.}
We first apply deterministic checks to repository metadata and the recorded Git tree.
These checks require an accessible source revision, a license permitting modification and redistribution, 10--4,999 GitHub stars, and a tracked Git tree of at most 500 MiB and 25,000 entries.

Projects that pass these checks are collected at the recorded revision with their README, setup documentation, package manifests, license files, and representative source paths.
We then use GPT-5.6-Luna to assess the project's setup path, evaluable process, runtime fit, and optimization potential from this source snapshot using a structured response schema.
The prompt is provided in Appendix~\ref{app:perfrace-filter-prompt}.

We require each structured response to address the following criteria.
\begin{list}{$\square$}{
  \setlength{\leftmargin}{1.6em}
  \setlength{\labelwidth}{1em}
  \setlength{\labelsep}{0.4em}
  \setlength{\topsep}{2pt}
  \setlength{\itemsep}{1pt}
  \setlength{\parsep}{0pt}
  \setlength{\partopsep}{0pt}
}
  \item \textbf{Setup path.} Repository files support concrete commands for installing dependencies and building or running the project.
  \item \textbf{Evaluable process.} The project supports a finite process with a fixed input, observable output, performance metric, and correctness check.
  \item \textbf{Runtime fit.} The process can run in a Linux x86-64 container on commodity CPUs without mandatory special hardware.
  \item \textbf{Optimization potential.} The project supports a substantial local computation whose cost is expected to increase with input size or repeated execution.
\end{list}

Using these assessments, we selected 239 high-potential project candidates for Pair Verification.
Table~\ref{tab:perfrace-filter-statistics} reports the first check that excludes each project from the final Pair Verification selection.

\input{tables/perfrace-filter-statistics}

\begin{figure}[H]
  \centering
  \includegraphics[width=\textwidth]{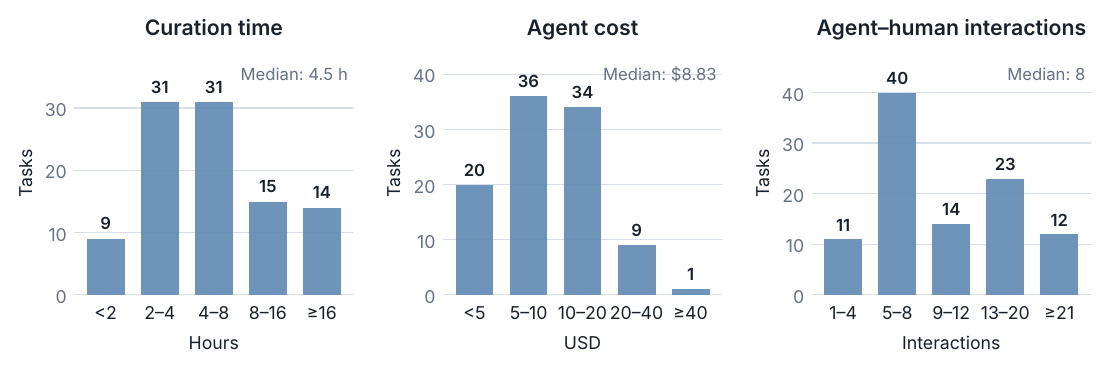}
  \caption{Distribution of curation time, agent cost, and agent--human interactions across the 100 accepted tasks.}
  \label{fig:perfrace-curation-effort}
\end{figure}

\textbf{Curation effort.}
To quantify the effort required to turn selected projects into accepted tasks, Figure~\ref{fig:perfrace-curation-effort} reports the agent cost, curation time, and number of agent--human interactions for the 100 accepted tasks.
The median task required 4.5 hours, \$8.83 in agent cost, and eight agent--human interactions, compared with averages of 9.1 hours, \$11.35, and 12 interactions.

\subsection{Pair Verification}
\label{app:perfrace-pair-verification}

\perfrace{} uses Pair Verification to ensure that each task represents meaningful project use, produces reproducible measurements, and rejects shortcut solutions; this section describes the agent--human protocol and analyzes the resulting curation trajectories.

\subsubsection{Protocol}

Pair Verification combines agent-led task construction with human review to turn each project candidate into an executable task and assess its realism, reliability, and integrity.
The construction skill is provided in Appendix~\ref{app:perfrace-pair-verification-skill}.
The agent develops the task in stages and pauses for a reviewer decision after completing the evidence required for each quality criterion.
\begin{figure}[H]
  \centering
  \includegraphics[width=\textwidth]{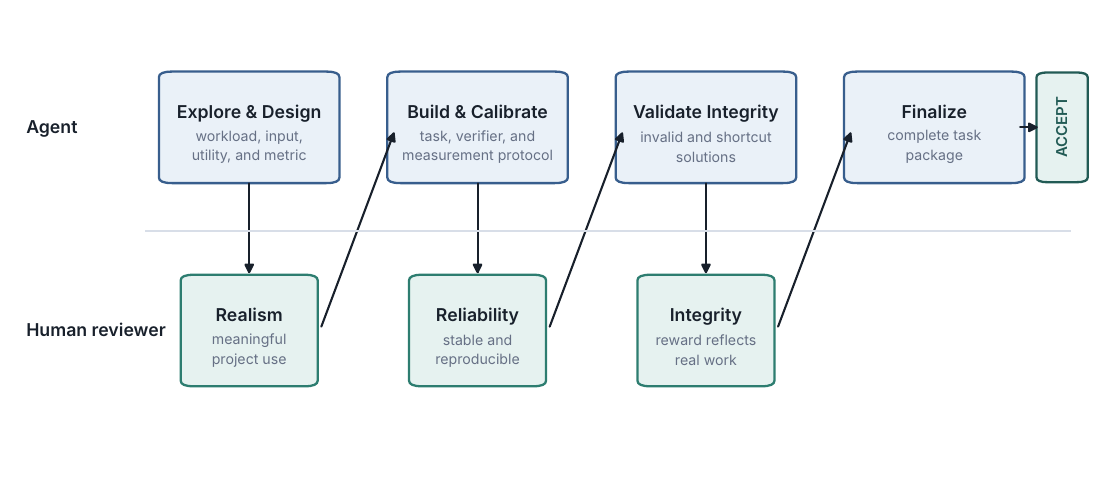}
  \caption{Pair Verification alternates between agent-led task construction and human review at the Realism, Reliability, and Integrity gates.}
  \label{fig:perfrace-pair-verification}
\end{figure}

\textbf{Explore and Design.}
The agent first studies the project's documentation, examples, code paths, and public usage evidence to understand its purpose, users, and performance-relevant operations.
The agent searches public release pages for source code, models, and assets, scientific or government data portals, downstream repositories, production archives, and real user exports to identify candidate workload inputs.
For each input, the agent records its download location, version, and applicable terms, and accepts it only when an explicit license, public-domain designation, or written authorization permits redistribution with the benchmark.

It uses these materials to identify candidate operations, but grounds the workload and scale in a real user or downstream application rather than repository-native tests, benchmarks, examples, or fixtures.
The agent compares plausible workload candidates, then traces the selected operation from its user-facing entry point through the relevant project code and demonstrates a small input-to-output execution on the unmodified revision.
During this run, it records dependency installation, project build, input preparation, and operation time to check that setup does not dominate the measurement and to estimate whether complete verification can finish within the 1,200-second budget.
The agent repeats the operation on identical inputs to detect run-to-run flakiness, including sporadic failures or inconsistent correctness-relevant outputs, and to identify any dependence of runtime on random state, network access, or external services.
A candidate is revised or rejected if its stochastic execution remains too variable for reproducible performance measurement after fixing available random seeds and aggregating repeated runs.

It then specifies the exact operation, workload and scale available to the agent, held-out correctness cases, observable correctness utility, performance metric, and measurement boundary.
Before requesting Realism review, the agent summarizes the project's purpose and users, the selected execution path, the basis for the input and scale, and the alternatives it considered and rejected.
The reviewer approves Realism only when the operation, input, and scale represent meaningful project use, the metric measures the stated utility, and the workload inputs can be redistributed as part of the benchmark.

\textbf{Build and Calibrate.}
After Realism approval, the agent derives the installation, build, and run commands from upstream documentation, CI configuration, and package manifests, and verifies them on the pinned source revision.
The agent implements the verifier, the task's test program that measures the performance metric on a separate copy of the workload and checks additional held-out cases for correctness.
It then constructs the Harbor task files with \texttt{instruction.md}, \texttt{task.toml}, an \texttt{environment/} directory containing the editable project and workload, and a separate \texttt{tests/} directory containing the verifier code, workload copy, held-out cases, and expected outputs.
The agent pins the project revision, base image, language toolchain, dependencies, workload assets, and expected outputs so that the image can be rebuilt and the task rerun without downloading project source or workload inputs.
The task instruction tells the agent how to build and run the project, what it may edit, which behavior it must preserve, and which performance metric to improve.

The agent then builds a fresh image and runs the verifier on the unmodified project before calibrating performance measurements.
The agent selects the smallest resource profile that completes the workload without CPU, memory, or storage pressure and sets the verifier timeout within 1,200 seconds.
The agent uses pilot and repeated runs to choose the workload scale, warmups, repetitions, and aggregation, requiring the final measurements to have a coefficient of variation below 10\% while keeping build and setup outside the timed operation.
The agent checks that the timing boundary covers the complete operation requested in the task while excluding one-time environment setup and verifier preparation that are unrelated to the optimization objective.

For the \(n\) timed measurements retained after warmup, \(x_1,\ldots,x_n\), we compute
\[
\mathrm{CV}=\frac{s}{\bar{x}},
\qquad
\bar{x}=\frac{1}{n}\sum_{i=1}^{n}x_i,
\qquad
s=\sqrt{\frac{1}{n-1}\sum_{i=1}^{n}(x_i-\bar{x})^2}.
\]

After fixing the workload, correctness checks, resources, and measurement protocol, the agent builds the final image and evaluates the unmodified baseline and one optimization candidate under the same conditions.
The reviewer examines the fresh-build correctness results, repeated measurements and CVs, verification time, resource allocation, and candidate trajectory.
The reviewer approves Reliability only when fresh builds and repeated runs return the same correctness verdict and stable performance measurements with the chosen warmups, repetitions, aggregation, and measurement boundary.

\textbf{Validate Integrity.}
To identify reward-hacking paths that could receive a valid score without completing the task, the agent tests deliberately incorrect, partial, hard-coded, and work-skipping solutions and checks that workload inputs, expected outputs, and verifier-owned data cannot be replaced by agent-controlled files.
The agent inspects the evaluation image for leaked expected outputs, verifier code, or construction notes that could reveal the hidden checks.
For example, in \texttt{rust-clippy--cargo-package-lint}, removing Cargo dependency tracking made later runs reuse a cached lint result, which the verifier detects by touching \texttt{clippy-driver}, running the workload twice against the same warm target, and requiring each Cargo JSON trace to report a non-fresh compiler artifact for \texttt{cargo}.

The reviewer approves Integrity only when every requirement in the task instruction is checked, every verifier check follows from the declared task, and none of the tested shortcut solutions can receive a valid score.

\textbf{Finalize.}
After all three reviews, the reviewer chooses \textsc{accept}, \textsc{revise}, or \textsc{reject}; accepted tasks pass final compatibility checks before publication as a task package comprising an immutable Docker image and task artifact.

\subsubsection{Quality Control}

We analyze the recorded agent--human Pair Verification trajectories for evidence of the Realism, Reliability, and Integrity of the resulting tasks.
Of the 239 candidates, 132 were rejected during Explore and Design, six during Build and Calibrate, none during Validate Integrity, and one during Finalize, leaving 100 accepted tasks.
\begin{figure}[H]
  \centering
  \includegraphics[width=0.78\textwidth]{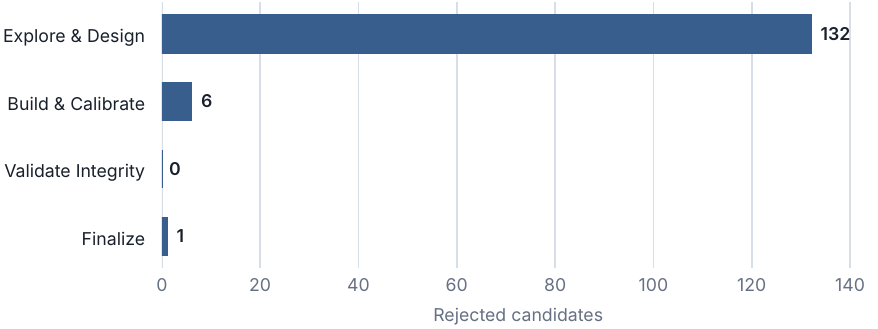}
  \caption{Candidates rejected at each Pair Verification stage, with candidates lacking a recorded transition into Build \& Calibrate assigned to Explore \& Design.}
  \label{fig:perfrace-stage-rejections}
\end{figure}
\textbf{Realism.}
Realism checks cover the workload operation and scale, the source and redistribution terms of its inputs, and the alignment between the declared utility and performance metric.
Table~\ref{tab:perfrace-workload-input-sources} groups the workload inputs into released source code, models, or assets (26 tasks), public scientific or government datasets (21), real downstream applications (21), production traces or archives (16), real user exports (6), and other independently grounded inputs (10).
\input{tables/perfrace-workload-input-sources}

\textbf{Reliability.}
All 99 wall-clock tasks have a final unmodified-project CV below 10\%; the remaining task, \texttt{metamorpho--reallocate}, measures transaction gas, the deterministic execution cost assigned by the Ethereum Virtual Machine to a fixed smart-contract transaction, and returns exactly 874,508 gas in three fresh containers.
Table~\ref{tab:perfrace-cv-distribution}(a) shows that 84 of the 99 wall-clock tasks have a final CV below 6\%.

\textbf{Integrity.}
We validate Integrity across six attack classes: empty or no-op solutions, skipped work, hard-coded outputs, precomputed or cached outputs, modification of protected inputs or tests, and access to verifier-only data.
All 100 accepted tasks completed an explicit red-team pass before Integrity approval.
Table~\ref{tab:perfrace-cv-distribution}(b) shows that 40 tasks were tested against at least two red-team cases, including two tasks with seven or eight cases.
For example, SCC is a source-code statistics tool whose task scans the Linux v6.19 tree and emits JSON statistics; during red-team testing, the agent replaced the scanner with a program that returned the precomputed Linux result instead of scanning the source tree, which passed the workload available to the agent but failed when the verifier asked it to scan a separate pinned checkout of SCC's own source code.
\input{tables/perfrace-cv-distribution}

\subsection{Evaluation Protocol}
\label{app:perfrace-evaluation}

For each submission, the Judge snapshots the agent's current container and starts a temporary shadow container from that snapshot.
The Judge creates the snapshot with \texttt{docker compose commit}, launches the shadow container under a separate Compose project, and removes its container, local image, and volumes after verification.
The agent container remains paused during verification and resumes from the same workspace state after the shadow container is removed.

Inside the shadow container, the verifier runs through \texttt{tests/test.sh} and emits the correctness verdict and performance metric as JSON.
The verifier code, workload copy, held-out cases, and expected outputs are absent from the agent container and available only inside the shadow container.
Each task declares its warmups, repetition count, aggregation rule, and required correctness pass rate in \texttt{task.toml}, and its verifier returns a performance metric only when that pass rate is met.

The task clock continues while the shadow is created, executed, and removed, so every submission trades part of budget \(B\) for additional correctness and performance feedback.
After verification, the Judge returns the correctness verdict and performance metric to the agent as feedback for subsequent optimization.
For a correct submission, the Judge computes the speedup score by dividing the unmodified baseline metric by the submission metric.
Submissions that fail correctness, exceed the verifier timeout, or do not produce a valid performance metric cannot update the best-so-far trajectory \(S(t)\).

\begin{figure}[H]
  \centering
  \includegraphics[width=0.82\textwidth]{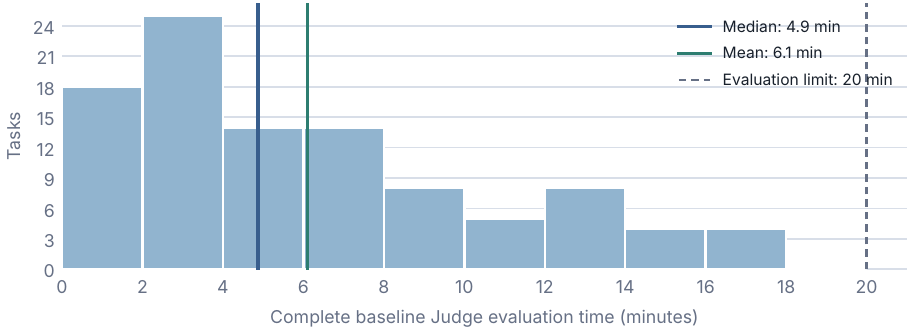}
  \caption{Distribution of complete baseline Judge evaluation time across the 100 final task packages, with a mean of 6.1 minutes, a median of 4.9 minutes, and a 20-minute evaluation limit.}
  \label{fig:perfrace-verification-time}
\end{figure}

Five minutes before the deadline, the agent receives a final delivery prompt asking it to leave the project in a coherent state before the Judge performs one final shadow verification of the current workspace.
Figure~\ref{fig:perfrace-verification-time} characterizes the time required for one complete Judge evaluation, showing that baseline Judge evaluation remains within the 20-minute limit for all 100 final tasks.

\phantomsection
\textbf{Longer optimization budgets.}
\label{app:perfrace-longer-budgets}
To measure the benefit of additional optimization time, we run Terminus-KIRA with GPT-5.6-Luna at high reasoning effort three times on each of the 88 evaluated tasks, maintaining one persistent agent session across Judge submissions during every four-hour run.

\begin{figure}[H]
  \centering
  \includegraphics[width=\textwidth]{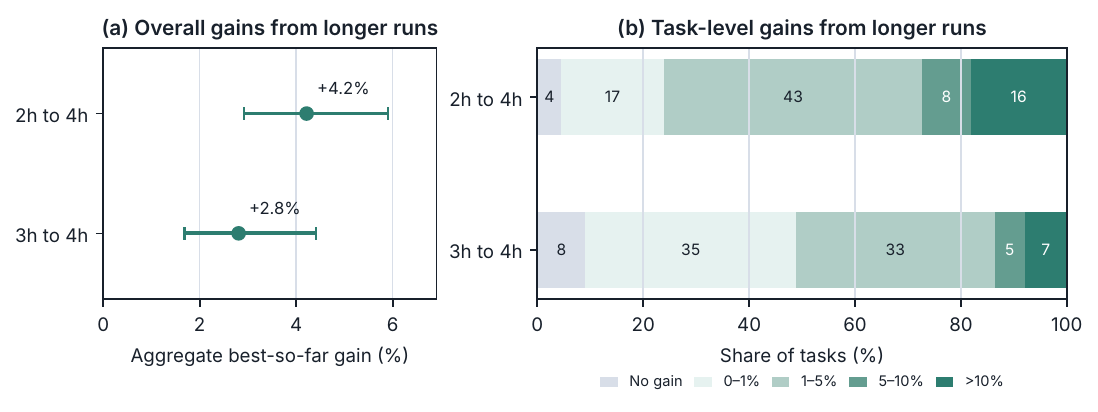}
  \caption{Performance improvements from additional optimization time, based on
  three four-hour runs for each of 88 \perfrace{} tasks. Left: increase in the overall
  score at four hours relative to two and three hours; horizontal lines show
  95\% bootstrap intervals. Right: number of tasks in each improvement range.}
  \label{fig:perfrace-late-budget-gains}
\end{figure}

The aggregate best-so-far correct speedup rises from \(1.314\times\) after one hour to \(1.377\times\), \(1.396\times\), and \(1.435\times\) after two, three, and four hours, respectively.
Across the resulting 264 trajectories, Figure~\ref{fig:perfrace-late-budget-gains} shows that extending
Terminus-KIRA's budget from two to four hours improves the overall score by
4.2\%, and the fourth hour alone adds 2.8\%.

The four-hour score is at least 5\% higher on 24 tasks than at two hours and on
12 tasks than at three hours, showing that agents continue to find meaningful
improvements late in the task.
Among the 228 runs that improve over baseline, the best result is first reached at a median of 88.1\% of the four-hour budget, and 51 of the 264 runs produce a new improvement of at least 5\% during the final hour.

%% file: tables/perfrace-comparison.tex
\begin{table}[H]
  \centering
  \scriptsize
  \setlength{\tabcolsep}{3pt}
  \renewcommand{\arraystretch}{1.25}
  \caption{Comparison with related benchmarks. Microbenchmarks measure targeted
  operations (for example, selected API calls), whereas end-to-end workloads
  measure complete project-use workflows.}
  \label{tab:perfrace-benchmark-comparison}
  \begin{tabularx}{\textwidth}{@{}l
    >{\hsize=0.9\hsize\linewidth=\hsize\raggedright\arraybackslash\hspace{0pt}}X
    >{\hsize=1.3\hsize\linewidth=\hsize\raggedright\arraybackslash\hspace{0pt}}X
    >{\hsize=0.8\hsize\linewidth=\hsize\raggedright\arraybackslash\hspace{0pt}}X@{}}
    \toprule
    Benchmark & Task type & Task construction & Evaluation \\
    \midrule
    \perfrace{} &
    Repository optimization on end-to-end workloads &
    Live GitHub discovery followed by agent--human Pair Verification &
    Iterative correctness and performance feedback from an external Judge \\
    AutoLab~\mbox{\citep{xu2026autolab}} &
    Diverse metric-driven optimization tasks &
    Expert-curated optimization tasks &
    Local evaluation during the task; held-out final evaluation \\
    EdgeBench~\mbox{\citep{zhu2026edgebench}} &
    Cross-domain iterative problem-solving tasks &
    Expert-curated day-scale tasks &
    Local inner loop; repeated submission-gated Judge feedback \\
    FrontierSWE~\mbox{\citep{chu2026frontierswe}} &
    Ultra-long-horizon software engineering and research tasks &
    Real-world problems curated with academic and industry experts &
    Final graded submission after up to 20 hours \\
    Terminal-Bench~\mbox{\citep{merrill2026terminalbench}} &
    General terminal tasks &
    Crowdsourced tasks with multi-stage human review &
    Final-only test-based evaluation \\
    GSO~\mbox{\citep{shetty2025gso}} &
    Microbenchmark-based repository optimization &
    Repository commit histories and generated performance tests &
    Final-only microbenchmark-based evaluation \\
    SWE-Perf~\mbox{\citep{he2025sweperf}} &
    Microbenchmark-based repository optimization &
    Historical performance PRs and selected repository tests &
    Final-only microbenchmark-based evaluation \\
    SWE-fficiency~\mbox{\citep{ma2025swefficiency}} &
    Microbenchmark-based repository optimization &
    Performance PRs and manually reconstructed workloads &
    Final-only microbenchmark-based evaluation \\
    FormulaCode~\mbox{\citep{sehgal2026formulacode}} &
    Microbenchmark-based repository optimization &
    Continuously ingested performance PRs and ASV workloads &
    Final-only microbenchmark-based evaluation \\
    \bottomrule
  \end{tabularx}
\end{table}

%% file: tables/perfrace-task-catalog.tex
{\scriptsize
\setlength{\tabcolsep}{3pt}
\renewcommand{\arraystretch}{1.12}
\begin{longtable}{@{}>{\raggedright\arraybackslash}p{0.22\textwidth}
  >{\raggedright\arraybackslash}p{0.23\textwidth}
  >{\raggedright\arraybackslash}p{0.13\textwidth}
  >{\raggedright\arraybackslash}p{0.36\textwidth}@{}}
  \caption{Complete catalog of the 100 \perfrace{} tasks.}
  \label{tab:perfrace-task-catalog}\\
  \toprule
  Task & Project and purpose & License & Fixed workload \\
  \midrule
  \endfirsthead
  \multicolumn{4}{c}{\tablename~\thetable{} (continued)} \\
  \toprule
  Task & Project and purpose & License & Fixed workload \\
  \midrule
  \endhead
  \midrule
  \multicolumn{4}{r}{Continued on next page} \\
  \endfoot
  \bottomrule
  \endlastfoot
  \path{aiidateam-aiida-core--battery-archive-import} & \path{aiidateam/aiida-core}\newline Scientific workflow and provenance platform & MIT & AiiDA import of a public battery-cycling provenance archive \\
  \path{anyrender--openmoji} & \path{DioxusLabs/anyrender}\newline Rust 2D drawing abstraction & Apache-2.0 & AnyRender batch rasterization of the OpenMoji SVG corpus \\
  \path{apalache--zcash-sync} & \path{apalache-mc/apalache}\newline Symbolic model checker for TLA+ and Quint & Apache-2.0 & Apalache's bounded safety check for the Zcash v2 block-download scheduler \\
  \path{arrow-rs--variant-wikidata} & \path{apache/arrow-rs}\newline Rust implementation of Apache Arrow & Apache-2.0 & apache/arrow-rs while encoding Wikidata entities as a Variant column \\
  \path{bevy-voxel-world--terrain-streaming} & \path{splashdust/bevy_voxel_world}\newline Voxel-world plugin for Bevy & Apache-2.0 & splashdust/bevy\_voxel\_world terrain generation and chunk meshing \\
  \path{bm25s--trec-covid-index} & \path{xhluca/bm25s}\newline BM25 search library & MIT & bm25s indexing and persistence for the TREC-COVID research collection \\
  \path{boyter-scc--linux-v6-19-json-scan} & \path{boyter/scc}\newline Source-code counting and complexity tool & MIT & SCC whole-repository JSON scan time on Linux v6.19 \\
  \path{calldiff--pi-review} & \path{tanishqkancharla/calldiff}\newline Function call-stack diff tool & MIT & Calldiff JSON call-stack diffs for a real Pi coding-agent review \\
  \path{cbcoutinho-nextcloud-mcp-server--ohr-pdf-ingest} & \path{cbcoutinho/nextcloud-mcp-server}\newline MCP server for Nextcloud & AGPL-3.0 & Nextcloud MCP Server's born-digital PDF ingest and page-aware chunking throughput \\
  \path{ccc--compile-speed} & \path{anthropics/claudes-c-compiler}\newline Dependency-free C compiler & CC0-1.0 & CCC compile-to-object latency on the SQLite 3.46.0 amalgamation \\
  \path{chaste--ttp06-sheet} & \path{Chaste/Chaste}\newline Biomedical simulation framework & BSD-3-Clause & Chaste cardiac simulation for a TTP06 ventricular sheet propagation \\
  \path{cmtraceopen--hadoop-logs} & \path{adamgell/cmtraceopen}\newline Cross-platform Windows log viewer & MIT & CMTrace Open while loading a real Hadoop/YARN log session \\
  \path{cocoon--oci-import} & \path{cocoonstack/cocoon}\newline MicroVM engine & MIT & local import of a supported Cocoon Ubuntu VM OCI image \\
  \path{codediff--neovim-release} & \path{esmuellert/codediff.nvim}\newline Neovim code-review and diff plugin & MIT & CodeDiff C diff latency for a real Neovim release review \\
  \path{commitizen--check} & \path{commitizen-tools/commitizen}\newline Commit and changelog automation tool & MIT & Commitizen changelog generation for Angular's complete Git history \\
  \path{core--blob-upload} & \path{calimero-network/core}\newline Peer-to-peer application platform & Apache-2.0 & Calimero content-addressed blob ingestion for a real OpenStreetMap file upload \\
  \path{cpp2rust--woff2-compress} & \path{Cpp2Rust/cpp2rust}\newline C++-to-safe-Rust translator & MIT & Cpp2Rust's safe Rust output for public-font WOFF2 compression \\
  \path{deepinv--kohler-kernels} & \path{deepinv/deepinv}\newline Deep-learning inverse-problems library & BSD-3-Clause & DeepInverse spatially varying blur-kernel estimation on a real Kohler camera-motion-blurred image \\
  \path{dftd4--lysozyme-gradient} & \path{dftd4/dftd4}\newline Atomic-charge-dependent dispersion correction & LGPL-3.0 & DFT-D4 energy, analytic gradient, and virial evaluation for a real lysozyme structure \\
  \path{dither-guy--photo-batch} & \path{manoelpiovesan/dither-guy}\newline Image dithering application & GPL-3.0 & Dither Guy's full-resolution photo batch export \\
  \path{doc-detective--redpanda-transforms} & \path{doc-detective/doc-detective}\newline Documentation testing framework & AGPL-3.0 & Doc Detective dry-run planning for Redpanda transform documentation \\
  \path{drasi-core--gtfs-delay-replay} & \path{drasi-project/drasi-core}\newline Continuous-query engine & Apache-2.0 & Drasi Core's continuous Cypher evaluation for replayed GTFS-Realtime vehicle geofence updates \\
  \path{duckdb--taxi-ingest} & \path{Giorgi/DuckDB.NET}\newline .NET bindings for DuckDB & MIT & DuckDB.NET ingestion of NYC Yellow Taxi trip records \\
  \path{dusty--fpindex-ingest} & \path{lalinsky/dusty}\newline Zig HTTP client/server library & MIT & Dusty's HTTP batch-ingestion path for real AcoustID fpindex updates \\
  \path{embree--chess-render} & \path{RenderKit/embree}\newline Ray-tracing kernels & Apache-2.0 & Embree triangle traversal for a public chess-scene visibility render \\
  \path{endee--esci} & \path{endee-io/endee}\newline Vector database & AGPL-3.0 & Endee HTTP dense search for a replay of Amazon Shopping Queries product searches \\
  \path{enzyme--mnist} & \path{EnzymeAD/Enzyme.jl}\newline Julia automatic-differentiation library & MIT & Enzyme.jl for Flux's CPU LeNet MNIST training epoch \\
  \path{eole--gpt2-wikitext-generate} & \path{eole-nlp/eole}\newline Language-modeling toolkit & MIT & EOLE CPU GPT-2 batch text generation on WikiText-103 prompts \\
  \path{eucalypt--config-render-speed} & \path{curvelogic/eucalypt}\newline Configuration language and interpreter & MIT OR Apache-2.0 & the eucalypt interpreter on a config-generation workload \\
  \path{eucalypt--eval-speed} & \path{curvelogic/eucalypt}\newline Configuration language and interpreter & MIT OR Apache-2.0 & the eucalypt interpreter on an Advent of Code workload \\
  \path{facet--venta-guild} & \path{Tim-Maes/Facet}\newline .NET DTO and mapping source generator & MIT & Facet package and source-generator compilation time for venta's Guild service \\
  \path{flexmeasures--smard-price-import} & \path{FlexMeasures/flexmeasures}\newline Energy-management platform & Apache-2.0 & FlexMeasures import of a full year of public SMARD day-ahead prices \\
  \path{fluidsynth-fluidsynth--bagatelles-midi-render} & \path{FluidSynth/fluidsynth}\newline SoundFont software synthesizer & LGPL-2.1 & FluidSynth offline rendering of a complete CC0 MIDI performance to WAV \\
  \path{geoparquet-io--hilbert-sort} & \path{geoparquet/geoparquet-io}\newline GeoParquet tooling & Apache-2.0 & GeoParquet Hilbert ordering of an Overture Buildings shard \\
  \path{gridflock--drawer-stl} & \path{yawkat/GridFlock}\newline Gridfinity baseplate generator & MIT / CC-BY-4.0 & GridFlock's segmented Gridfinity drawer-baseplate STL export \\
  \path{ifcopenshell--gni-bimproject-obj-export} & \path{IfcOpenShell/IfcOpenShell}\newline IFC library and geometry engine & LGPL-3.0 & IfcOpenShell IfcConvert OBJ export for a lighter GNI BIM Project architectural IFC delivery \\
  \path{intro-skipper--credits} & \path{intro-skipper/intro-skipper}\newline Jellyfin intro and credits detector & GPL-3.0 & Intro Skipper's movie credits black-frame evidence extraction \\
  \path{jagua--nesting} & \path{JeroenGar/jagua-rs}\newline 2D collision engine for nesting & MPL-2.0 & jagua-rs collision collection replayed from a downstream nesting solver \\
  \path{javaevolved--site-build} & \path{javaevolved/javaevolved.github.io}\newline Java developer documentation site & MIT & the full localized GitHub Pages build for java.evolved \\
  \path{jsonnet--kubernetes-mixin-speed} & \path{google/jsonnet}\newline Data templating language & Apache-2.0 & the Jsonnet C++ interpreter on a Kubernetes monitoring configuration workload \\
  \path{juliaio-zarr-jl--oceananigans-timeseries-output} & \path{JuliaIO/Zarr.jl}\newline Julia implementation of Zarr array storage & MIT & Zarr.jl directory-store output for a real Oceananigans tracer time series \\
  \path{julialang-julia--trixi-jeans-instability} & \path{JuliaLang/julia}\newline Julia programming language & MIT & Julia runtime execution for a published Trixi Jeans gravitational-instability simulation \\
  \path{kitops--minilm-pack} & \path{kitops-ml/kitops}\newline OCI packaging tool for ML artifacts & Apache-2.0 & KitOps local ModelKit packaging for a complete all-MiniLM-L6-v2 release snapshot \\
  \path{kiwifs--kubernetes-reindex} & \path{kiwifs/kiwifs}\newline Markdown filesystem for agents and teams & BUSL-1.1 & KiwiFS SQLite FTS5 reindex for Kubernetes documentation \\
  \path{kms--batch-encrypt} & \path{Cosmian/kms}\newline Key-management system & BUSL-1.1 & Eviden KMS KMIP BulkData AES-GCM encryption for concurrent application-record batches \\
  \path{kubb--esi} & \path{kubb-labs/kubb}\newline OpenAPI code-generation framework & MIT & Kubb regeneration of JitaSpace's EVE Online ESI client \\
  \path{latexsnipper--formula-batch} & \path{SakuraMathcraft/LaTeXSnipper}\newline Formula OCR and editing tool & GPL-3.0 & resident CPU formula OCR for a public OpenStax textbook screenshot batch \\
  \path{libcachesim--twitter-lru} & \path{1a1a11a/libCacheSim}\newline Cache simulation library & Apache-2.0 & Four-capacity LRU cache-size sweep over a production Twitter request trace \\
  \path{lsp-tree-sitter--termux-check} & \path{neomutt/lsp-tree-sitter}\newline Tree-sitter language-server library & GPL-3.0 & Tree-sitter validation of a real Termux package-recipe corpus \\
  \path{mai-gen-videob50--score-images} & \path{Nick-bit233/mai-gen-videob50}\newline Maimai score-video generation tool & MIT & batch maimai score-image rendering for a real 50-record archive \\
  \path{meshmonitor--coverage-grid} & \path{Yeraze/meshmonitor}\newline Mesh-node monitoring web tool & BSD-3-Clause & MeshMonitor's SQLite coverage-grid batch \\
  \path{metamorpho--reallocate} & \path{morpho-org/metamorpho}\newline Tokenized lending-vault protocol & GPL-2.0 & Gas use for a historical MetaMorpho market reallocation \\
  \path{mini-qr--library-batch} & \path{lyqht/mini-qr}\newline QR-code generator and scanner & GPL-3.0 & Mini QR's batch SVG export for a public library directory \\
  \path{mirage--kubernetes-triage} & \path{strukto-ai/mirage}\newline Virtual terminal for coding agents & Apache-2.0 & Mirage recursive source search for Kubernetes dual-stack readiness-probe triage \\
  \path{modflow6--neversink} & \path{MODFLOW-ORG/modflow6}\newline Groundwater-flow simulator & CC0-1.0 & Complete steady-state simulation of the USGS Neversink--Rondout groundwater model \\
  \path{monorepo--crypto-build} & \path{commonwarexyz/monorepo}\newline Distributed-systems primitives & Apache-2.0 & the release build latency of commonware-cryptography \\
  \path{mrst--norne-import} & \path{SINTEF-AppliedCompSci/MRST}\newline Reservoir-simulation toolbox & GPL-3.0 & MRST Norne field-model import and geometry construction \\
  \path{msdialworkbench--lcms-alignment} & \path{systemsomicslab/MsdialWorkbench}\newline Mass-spectrometry analysis workbench & LGPL-3.0 & MS-DIAL Console LC-MS peak processing and alignment for a real negative-ion DDA batch \\
  \path{mssql-extension--taxi-copy} & \path{hugr-lab/mssql-extension}\newline DuckDB extension for Microsoft SQL Server & MIT & BCP bulk import of NYC taxi records through the DuckDB MSSQL extension \\
  \path{n8n-mcp--tool-batch-speed} & \path{czlonkowski/n8n-mcp}\newline MCP server for n8n workflow construction & MIT & n8n-mcp on a fixed MCP tool-call batch \\
  \path{nerdbank-messagepack--earthquake-fanout} & \path{AArnott/Nerdbank.MessagePack}\newline .NET MessagePack library & MIT & SignalR MessagePack earthquake snapshot fan-out \\
  \path{ojp--taxi-export} & \path{Open-J-Proxy/ojp}\newline Database proxy and JDBC driver & Apache-2.0 & OJP JDBC export of public NYC TLC taxi records \\
  \path{omnigraph--metabolic-trials-refresh} & \path{ModernRelay/omnigraph}\newline Lakehouse-native graph engine & MIT & OmniGraph's paginated metabolic clinical-trial refresh and lifecycle dashboard \\
  \path{onnxscript--bert-optimize} & \path{microsoft/onnxscript}\newline ONNX model authoring and optimization library & MIT & Graph optimization of a BERT-SQuAD FP32 and INT8 model bundle \\
  \path{oxidize-pdf--rag-chunking} & \path{bzsanti/oxidizePdf}\newline Rust PDF library for RAG & MIT & oxidize-pdf RAG chunking over a real CC-BY PDF corpus \\
  \path{pdal--dtm} & \path{PDAL/PDAL}\newline Point-cloud processing library & BSD-3-Clause & Ground classification and 1 m terrain-model generation from USGS 3DEP LiDAR \\
  \path{petsc--ocean-age-solve} & \path{JuliaParallel/PETSc.jl}\newline Julia wrappers for PETSc & MIT & PETSc.jl GMRES+ILU for a public OCCA ocean-age transport solve \\
  \path{physical-ai-toolchain--action-smoothness} & \path{microsoft/physical-ai-toolchain}\newline Physical-AI dataset analysis toolkit & MIT & multi-dataset LeRobot action-smoothness analysis for real ALOHA manipulation recordings \\
  \path{plonky3--ethereum-calldata-keccak-proof} & \path{Plonky3/Plonky3}\newline Polynomial IOP toolkit & Apache-2.0 & Plonky3 Ethereum calldata Keccak trace generation and STARK proving \\
  \path{polkadot-api--descriptor-generation} & \path{polkadot-api/polkadot-api}\newline Modular Polkadot API libraries & MIT & PAPI's multi-chain TypeScript descriptor generation \\
  \path{pytrendy--dow-index} & \path{RussellSB/pytrendy}\newline Trend-detection library & MIT & PyTrendy trend detection for a real daily Dow Jones Industrial Average price series \\
  \path{qrisp--modmul-compile} & \path{eclipse-qrisp/Qrisp}\newline Quantum-algorithm development framework & EPL-2.0 & Qrisp compilation of controlled modular multiplication \\
  \path{react-native-code-push--mattermost-release} & \path{Soomgo-Mobile/react-native-code-push}\newline React Native OTA update client & MIT & CodePush Android OTA releases for Mattermost Mobile \\
  \path{reactant--fashion-mnist-train} & \path{EnzymeAD/Reactant.jl}\newline MLIR/XLA compiler for Julia & MIT & Reactant CPU execution of a real Lux Fashion-MNIST training epoch \\
  \path{rspress--react-native-paper} & \path{web-infra-dev/rspress}\newline Static-site generator & MIT & Rspress production builds of the React Native Paper documentation site \\
  \path{rust-clippy--cargo-package-lint} & \path{rust-lang/rust-clippy}\newline Rust linter & Apache-2.0 & Clippy lint time for Cargo's main production package \\
  \path{s3s--kubernetes-sync} & \path{s3s-project/s3s}\newline S3 service adapter & Apache-2.0 & s3s-fs recursive S3 synchronization of Kubernetes pkg source \\
  \path{senpi--image-batch} & \path{code-yeongyu/senpi}\newline Coding-agent runtime & MIT & Senpi image attachment preprocessing for a real screenshot batch \\
  \path{snarkos--cdn-replay} & \path{ProvableHQ/snarkOS}\newline Decentralized operating system for ZK applications & Apache-2.0 & snarkOS CDN block bundle replay over 150 fixed Aleo Mainnet blocks \\
  \path{snarkvm--mainnet-execution-verify} & \path{ProvableHQ/snarkVM}\newline Zero-knowledge virtual machine & Apache-2.0 & snarkVM verification of a batch of accepted Aleo mainnet executions \\
  \path{solvespace-solvespace--printer-assembly-mesh-export} & \path{solvespace/solvespace}\newline Parametric 2D/3D CAD & GPL-3.0 & SolveSpace CLI mesh export for a complete Cartesian 3D-printer assembly \\
  \path{spectral--api-lint} & \path{stoplightio/spectral}\newline OpenAPI, JSON, and YAML linter & Apache-2.0 & Spectral linting of a production OpenAPI batch \\
  \path{sqruff--glam-etl} & \path{quarylabs/sqruff}\newline SQL formatter and linter & Apache-2.0 & Sqruff directory linting for Mozilla GLAM ETL SQL \\
  \path{strawberry-graphql-strawberry--graphique-tlc-weekly-columns} & \path{strawberry-graphql/strawberry}\newline Python GraphQL library & MIT & Strawberry GraphQL execution for Graphique's real TLC weekly Parquet column query \\
  \path{suitesparse--livejournal-pagerank} & \path{DrTimothyAldenDavis/SuiteSparse}\newline Sparse-matrix and graph computation suite & Apache-2.0 / BSD-2-Clause (used components) & Complete PageRank computation on the LiveJournal social graph using GraphBLAS and LAGraph \\
  \path{sum--text-attest} & \path{OtotaO/SUM}\newline Cross-runtime text-attestation library & Apache-2.0 & SUM while batch-attesting a corpus of real Wikipedia prose \\
  \path{sundials--cantera-ignition} & \path{llnl/sundials}\newline Nonlinear and differential-equation solver suite & BSD-3-Clause & SUNDIALS CVODES for a downstream Cantera methane ignition parameter sweep \\
  \path{svf-tools-svf--nginx-andersen-wpa} & \path{SVF-tools/SVF}\newline Static value-flow analysis framework & AGPL-3.0-or-later & SVF Andersen whole-program pointer analysis on NGINX reverse-proxy bitcode \\
  \path{synalinks--kubernetes-tasks} & \path{SynaLinks/synalinks}\newline Neuro-symbolic language-model framework & Apache-2.0 & Synalinks Markdown knowledge-base ingestion for Kubernetes task documentation \\
  \path{tfhe-rs--erc7984-transfer} & \path{zama-ai/tfhe-rs}\newline Fully homomorphic encryption library & BSD-3-Clause-Clear & TFHE-rs CPU execution for confidential ERC-7984 token transfers \\
  \path{tileserver-gl--country-atlas} & \path{maptiler/tileserver-gl}\newline Vector and raster map-tile server & BSD-2-Clause & TileServer GL static PNG rendering for a full Natural Earth country atlas \\
  \path{trixi-framework-trixi-jl--self-gravitating-sedov-amr} & \path{trixi-framework/Trixi.jl}\newline Adaptive numerical-simulation framework & MIT & a published self-gravitating Sedov blast-wave simulation with adaptive mesh refinement \\
  \path{typst--napkin} & \path{Myriad-Dreamin/typst.ts}\newline Typst runtime for JavaScript & Apache-2.0 & Typst.ts rendering in the Napkin Community production static-site build \\
  \path{vmecpp--precise-qh} & \path{proximafusion/vmecpp}\newline Magnetohydrodynamic equilibrium solver & MIT & VMEC++ fixed-boundary precise-QH equilibrium solves \\
  \path{wiredtiger--pageviews} & \path{wiredtiger/wiredtiger}\newline Embedded storage engine & GPL-2.0 OR GPL-3.0 & WiredTiger's durable import of a real Wikimedia pageview hour \\
  \path{xet-core--bert-download} & \path{huggingface/xet-core}\newline Content-addressed transfer and reconstruction engine & Apache-2.0 & local Xet reconstruction of a production BERT model artifact \\
  \path{yamato-security-hayabusa--external-attack-evtx-timeline} & \path{Yamato-Security/hayabusa}\newline Windows event-log forensics tool & AGPL-3.0 & Hayabusa bulk external attack-EVTX JSONL timeline generation \\
  \path{youtrackdb--osm-roads} & \path{JetBrains/youtrackdb}\newline Object-oriented graph database & Apache-2.0 & embedded YouTrackDB import of a real OpenStreetMap road network \\
  \path{zenbu--release-source} & \path{zenbu-labs/zenbu.js}\newline JavaScript application framework & MIT & Zenbu.js staging of the Zenbu desktop application's release source \\
  \path{zio--zrk-typecheck} & \path{lalinsky/zio}\newline Zig asynchronous-I/O framework & MIT & ZIO's downstream zrk ReleaseFast type-check latency \\
\end{longtable}
}

%% file: tables/perfrace-discovery-statistics.tex
\begin{table}[H]
  \centering
  \small
  \setlength{\tabcolsep}{10pt}
  \renewcommand{\arraystretch}{1.15}
  \caption{GitHub activity signals used for Discovery.}
  \label{tab:perfrace-discovery-statistics}
  \begin{tabular}{@{}lr@{}}
    \toprule
    Signal & Repositories \\
    \midrule
    Branch push by a coding-agent account & 112,127 \\
    Pull request opened by a coding-agent account & 6,390 \\
    \bottomrule
  \end{tabular}
\end{table}

%% file: tables/perfrace-filter-statistics.tex
\begin{table}[H]
  \centering
  \small
  \setlength{\tabcolsep}{10pt}
  \renewcommand{\arraystretch}{1.15}
  \caption{First failed check during Filter and the final selection result.}
  \label{tab:perfrace-filter-statistics}
  \begin{tabular}{@{}lr@{}}
    \toprule
    Result & Projects \\
    \midrule
    Fewer than 10 GitHub stars & 93,992 \\
    Accessibility & 1,134 \\
    Setup path & 212 \\
    Runtime fit & 359 \\
    Optimization potential & 4,462 \\
    Evaluable process & 3,066 \\
    Assessment error & 11 \\
    \midrule
    Selected for Pair Verification & 239 \\
    \bottomrule
  \end{tabular}
\end{table}

%% file: tables/perfrace-workload-input-sources.tex
\begin{table}[H]
  \centering
  \small
  \setlength{\tabcolsep}{6pt}
  \renewcommand{\arraystretch}{1.15}
  \caption{Sources of workload inputs across the 100 accepted tasks.}
  \label{tab:perfrace-workload-input-sources}
  \begin{tabularx}{\textwidth}{@{}>{\raggedright\arraybackslash}p{0.32\textwidth}
                                      >{\raggedright\arraybackslash}X
                                      r@{}}
    \toprule
    Input source & Examples & Tasks \\
    \midrule
    Released source code, model, or asset
      & Linux source tree, OpenMoji assets, released ML models & 26 \\
    Public scientific or government dataset
      & TREC-COVID, NYC taxi records, LC-MS data & 21 \\
    Real downstream application
      & React Native Paper, Headlamp, Mattermost Mobile & 21 \\
    Production trace or archive
      & Hadoop logs, GTFS delay feeds, GitHub events & 16 \\
    Real user export
      & MIDI collection, BIM project, printer assembly CAD & 6 \\
    Other independently grounded input
      & Wikidata variants, Dow index, OSM roads & 10 \\
    \bottomrule
  \end{tabularx}
\end{table}

%% file: tables/perfrace-cv-distribution.tex
\begin{table}[H]
  \centering
  \small
  \setlength{\tabcolsep}{7pt}
  \renewcommand{\arraystretch}{1.15}
  \begin{minipage}[t]{0.34\textwidth}
    \vspace{0pt}
    \centering
    \textbf{(a) Final measurement CVs}\par\smallskip
    \begin{tabular}{@{}lr@{}}
      \toprule
      CV range & Tasks \\
      \midrule
      \(0\% \leq \mathrm{CV} < 2\%\) & 20 \\
      \(2\% \leq \mathrm{CV} < 6\%\) & 64 \\
      \(6\% \leq \mathrm{CV} < 7\%\) & 8 \\
      \(7\% \leq \mathrm{CV} < 8\%\) & 3 \\
      \(8\% \leq \mathrm{CV} < 9\%\) & 4 \\
      \(9\% \leq \mathrm{CV} < 10\%\) & 0 \\
      \bottomrule
    \end{tabular}
  \end{minipage}
  \hspace{0.035\textwidth}
  \begin{minipage}[t]{0.27\textwidth}
    \vspace{0pt}
    \centering
    \textbf{(b) Red-team tests}\par\smallskip
    \begin{tabular}{@{}rr@{}}
      \toprule
      Tests per task & Tasks \\
      \midrule
      1 & 60 \\
      2 & 20 \\
      3 & 12 \\
      4 & 6 \\
      7 & 1 \\
      8 & 1 \\
      \bottomrule
    \end{tabular}
  \end{minipage}
  \caption{Reliability and integrity evidence: final CVs for the 99 wall-clock tasks and red-team tests for all 100 tasks, with one test assigned conservatively when an exact count cannot be recovered.}
  \label{tab:perfrace-cv-distribution}
\end{table}

%% file: src/8c_boxbox_appendix.tex
\section{\boxbox{}: Method Details}
\label{app:boxbox}

Figure~\ref{fig:boxbox-pseudocode} summarizes the task and review loops.
The following explains how harness revisions affect subsequent interactions
and how review, hot reloading, and recovery preserve task continuity.

\begin{figure}[H]
  \centering
\begingroup
\small
\definecolor{heading}{HTML}{385E8D}
\setlength{\parindent}{0pt}
\newenvironment{pseudocode}{%
  \begin{list}{\color{gray}\scriptsize\arabic{enumi}}{%
    \usecounter{enumi}
    \setlength{\leftmargin}{1.7em}
    \setlength{\labelwidth}{1.2em}
    \setlength{\labelsep}{.5em}
    \setlength{\itemsep}{1.5pt}
    \setlength{\parsep}{0pt}
    \setlength{\topsep}{4pt}}%
}{\end{list}}
\newcommand{\Line}[2]{\item\hspace*{#1em}#2}
\newcommand{\kw}[1]{\textbf{#1}}
\begin{minipage}{\linewidth}
{\normalsize\bfseries\color{heading} \boxbox{}}
\par\vspace{5pt}\hrule\vspace{6pt}
\textbf{State:} harness $H$, runtime $R$, task environment $e$, task trajectory $\tau$.
\par\textbf{Review state:} environment $e_r$ and separate trajectory $\tau_r$.
\par\textbf{Budget:} shared task budget $B$.

\begin{pseudocode}
  \Line{0}{\kw{while} budget $B$ remains \kw{do}}
  \Line{1}{$a \sim p_{\mathcal{M},H}(\cdot\mid\tau)$}
  \Line{1}{\kw{if} $a=\texttt{boxbox()}$ \kw{then}}
  \Line{2}{Yield control to the kernel; suspend task execution and preserve context}
  \Line{2}{Prepare $(e_r,\tau_r)$ for independent review under $H$, with access to $\tau$}
  \Line{2}{\kw{while true do}}
  \Line{3}{$a_r \sim p_{\mathcal{M},H}(\cdot\mid\tau_r)$}
  \Line{3}{\kw{if} $a_r=\texttt{commit}(H')$ \kw{then break}}
  \Line{3}{$(e_r,o_r) \gets \operatorname{Execute}_H(e_r,a_r)$}
  \Line{3}{$\tau_r \gets \tau_r\oplus(a_r,o_r)$}
  \Line{2}{\kw{end while}}
  \Line{2}{\kw{if} $H'\ne H$ \kw{then}}
  \Line{3}{The kernel validates $H'$ and attempts activation of its runtime $R'$}
  \Line{3}{\kw{if} successful \kw{then} select $(H,R)\gets(H',R')$}
  \Line{3}{\kw{else} retain or restore the previous harness and runtime}
  \Line{2}{\kw{end if}}
  \Line{2}{Restore continuation context and return control to the selected runtime}
  \Line{1}{\kw{else}}
  \Line{2}{$(e,o) \gets \operatorname{Execute}_H(e,a)$}
  \Line{2}{$\tau \gets \tau\oplus(a,o)$}
  \Line{1}{\kw{end if}}
  \Line{0}{\kw{end while}}
\end{pseudocode}
\vspace{4pt}\hrule
\end{minipage}
\endgroup
  \caption{\textbf{\boxbox{} pseudocode.} Task execution and independent harness review form nested interaction loops, with kernel-managed hot reloading connecting them.}
  \label{fig:boxbox-pseudocode}
\end{figure}

\subsection{Effects on Subsequent Interactions}
\label{app:boxbox-update-effects}

Equation~\ref{eq:boxbox-interaction} describes two ways a harness revision
can affect subsequent execution while the language model $\mathcal{M}$ remains
fixed. Changes to how the harness constructs model inputs can alter
$p_{\mathcal{M},H_k}$, influencing the agent's next action. Changes to
$\operatorname{Execute}_{H_k}$ can alter how that action is carried out and how
its result is returned as an observation. A revision can affect either or
both, so its influence extends to the trajectory generated after activation.

The current harness also governs the review loop. Once a revision is
activated, subsequent reviews use the updated implementation and can examine
the experience accumulated under it. Harness changes can therefore affect
both task execution and the process used to produce later revisions.
Appendix~\ref{app:case-study} illustrates concrete changes to guidance,
measurement, command polling, and context handling.

\subsection{Review}
\label{app:boxbox-agent-initiated}

The task-solving agent requests review through \texttt{boxbox()} without
having to propose a patch. The initial harness requires intervening task
activity and a 30-minute cooldown after each completed review; once eligible,
the agent decides whether and when to request another review.
The kernel preserves the task conversation and launches a separate agent
review session under the current harness. The reviewer can inspect task
trajectories and Git history, but edits only the harness, treating project
code as read-only. It may retain the current harness, add guidance, or modify
executable behavior, committing any changes before ending review.
The default review skill is provided in Appendix~\ref{app:boxbox-review-skill}.

\subsection{Execution Context}
\label{app:boxbox-execution-context}

Execution context includes both runtime-owned state and the existing task
environment. The former includes the task conversation, review history, and
records of which trajectory entries have already been reviewed. These are
exported as a state snapshot. The terminal session and project workspace
remain in place: shell state, ongoing processes, and project files are
preserved through the existing environment connection.

When the agent calls \texttt{boxbox()}, the runtime yields control to the
kernel, suspending the task interaction loop while review proceeds separately.
This pauses the agent's task decisions, not the entire container or its
processes. Before importing state into a candidate runtime, the kernel makes
a separate copy, preserving the original snapshot for recovery if the
candidate fails. This separates changing the agent implementation from
restarting its task environment.

\subsection{Hot Reloading}
\label{app:boxbox-transition}

Harness revisions are tracked as Git commits, while the kernel remains
separate from the editable harness. Before switching, the kernel checks the
committed candidate, loads its runtime, and verifies that it supports task
execution and context transfer. The current runtime keeps its terminal
connection until the candidate successfully imports the copied state.
The kernel then transfers the same terminal session to the candidate and
returns control to it, resuming task interaction from the preserved
conversation in the same workspace. If activation fails, the kernel retains
or restores the previous runtime and context.
For agent-requested reviews, the reviewer can leave concise guidance on using
new capabilities and continuing the task. This guidance is preserved across
the update, and the resumed agent is prompted to read it before continuing.
Task execution and review share the same budget and cost accounting.
Table~\ref{tab:boxbox-continuation} summarizes how execution continues under
each outcome.

\begin{table}[htbp]
  \centering
  \small
  \caption{Task continuation across review and runtime-update outcomes.}
  \label{tab:boxbox-continuation}
  \renewcommand{\arraystretch}{1.15}
  \begin{tabularx}{\linewidth}{@{}>{\raggedright\arraybackslash}p{.26\linewidth}X@{}}
    \toprule
    Outcome & Continuation \\
    \midrule
    No harness change &
    Resume the current runtime with the preserved task context. \\
    Successful update &
    Continue under the revised harness with transferred context, the existing
    terminal session, and the same project workspace. \\
    Update fails &
    Retain or restore the previous runtime and its preserved context. \\
    Runtime fault &
    Enter recovery and review using the retained execution records; restart
    the terminal while preserving project files. \\
    \bottomrule
  \end{tabularx}
\end{table}

\subsection{Recovery}
\label{app:boxbox-validation}

The kernel retains a recovery point across harness updates. A newly activated
runtime remains provisional until it reaches the next agent-requested review
without a runtime fault. If it fails before that boundary, the kernel restores
the previous confirmed harness and state and initiates a review using the
fault evidence. Faults in a confirmed runtime also trigger review.
Runtime-fault recovery starts a fresh terminal, so shell-local state is lost,
but project files, harness history, and execution records remain available.
The retained records let subsequent reviews examine failed revisions and help
the task-solving agent determine how to continue interrupted work.

%% file: src/8d_prompt_and_skill.tex
\section{Prompts and Skills}
\label{app:prompts-and-skills}

This appendix presents the Filter prompt and Pair Verification skill used in
benchmark construction, followed by the default \boxbox{} self-review skill.

\tcbset{paperprompt/.style={
  enhanced,
  breakable,
  listing only,
  fonttitle=\small\bfseries,
  center title,
  colback=white,
  colframe=black!18,
  colbacktitle=black,
  coltitle=white,
  boxrule=.35pt,
  arc=2pt,
  left=7pt,
  right=7pt,
  top=6pt,
  bottom=6pt,
  toptitle=2pt,
  bottomtitle=2pt,
  before skip=8pt,
  after skip=8pt,
  listing options={
  language={},
  literate={≤}{{$\leq$}}1 {—}{{---}}3,
  basicstyle=\ttfamily\fontsize{8.5}{10}\selectfont,
  columns=fullflexible,
  keepspaces=true,
  breaklines=true,
  breakatwhitespace=true,
  breakautoindent=false,
  breakindent=0pt,
  showstringspaces=false,
  numbers=none,
  frame=none,
  xleftmargin=0pt,
  xrightmargin=0pt,
  aboveskip=0pt,
  belowskip=0pt,
  linebackgroundcolor={}
  }
}}

\subsection{Filter Prompt}
\label{app:perfrace-filter-prompt}

The Filter assesses repository snapshots before Pair Verification.

% Source: perfbench/collect/filter/prompt.md at ec4637de510fcc64ddee1a94348bac3a8f682c7c.
% Condensed and reworded from the source; gate criteria retained, detailed carrier taxonomy omitted.
\begin{tcblisting}{paperprompt,
  title={Filter Prompt},
  title after break={Filter Prompt \normalfont\itshape (continued)}
}
You are the automated Filter for PerfRace project candidates. Decide whether the supplied repository snapshot provides enough source-grounded evidence to proceed to Pair Verification.

Evaluate four gates:

1. Accessible. The repository and revision are available, with an explicit license permitting modification and redistribution. A fork must have an independently maintained product identity.

2. Upstream reproducibility. Documentation, scripts, or manifests support concrete setup/build and minimal smoke commands. Cite the supporting path or manifest field for each command. Dependencies and configuration must come from identifiable public sources.

3. Evaluation potential. Identify a finite project operation with fixed inputs, clear start and end, observable output, a plausible efficiency metric, and a correctness check. Establish a replayable real user or downstream workload, or a self-contained computation with grounded input semantics. Repository tests, benchmarks, demos, fixtures, and generated inputs do not establish a portable workload. Flag reliance on unavailable deployment state, private traces, physical devices, or an unresolved workload source.

4. Resource feasibility. The operation must run in a Linux x86-64 container on commodity CPUs within ordinary resource limits. Reject mandatory special hardware, unavailable private services, and thin remote pass-throughs. Local services or fixed stubs are acceptable only when meaningful project-controlled computation remains.

Evidence and optimization potential:
- Use only the supplied snapshot; cite metadata or repository paths for every gate. Missing evidence fails the gate.
- Describe a preliminary process only when evaluation potential passes.
- Describe the local computation and its strongest optimization opportunity. Rate potential as low for tiny or externally dominated work, medium for a clear local hot path, and high for scale-sensitive computation with multiple structural opportunities.
- Return concise evidence in the requested structured response.
\end{tcblisting}

\subsection{Pair Verification Skill}
\label{app:perfrace-pair-verification-skill}

The skill guides agent-led construction through human reviews of Realism,
Reliability, and Integrity, followed by task publication.

% Source: perfbench/curate/skill/SKILL.md at ec4637de510fcc64ddee1a94348bac3a8f682c7c.
% Condensed and reworded from the source; review gates and key constraints retained.
% Directory layout, command syntax, reporting conventions, and detailed operational procedures omitted.
\begin{tcblisting}{paperprompt,
  title={Pair Verification Skill},
  title after break={Pair Verification Skill \normalfont\itshape (continued)}
}
# PerfRace Pair Verification

Turn one repository into a machine-evaluable task. The agent investigates and executes; the human reviews Realism, Reliability, and Integrity, then decides ACCEPT, REVISE, or REJECT.

## Working agreement

Maintain a report linking claims to project sources, runnable probes, and raw results. At each review gate, present the evidence, uncertainty, and recommendation, allow discussion, and pause for the human decision. Reopen the affected reviews when later evidence invalidates an approved claim.

## Stage 1: Explore & Design

Understand the project's users and performance-sensitive workflows. Compare candidate workloads and select an independently grounded real user or downstream operation with reproducible inputs and scale. Upstream examples and benchmarks may guide discovery but must not become the scored workload.

Run the unmodified workload end to end and identify the project execution path. Estimate build cost, image size, verification time, and measurement variance before constructing the task. Propose the operation, input provenance and scale, visible workload, independent verifier copy, additional correctness cases, correctness utility, measurement boundary, and metric.

Human review: Realism. Explain why the operation, inputs, and scale represent meaningful project use, which alternatives were considered, and what remains uncertain. Obtain approval before building the canonical task.

## Stage 2: Build & Calibrate

Package the pinned project, visible workload, instructions, container, and verifier. Keep only the project writable. Use an independent byte-identical scored input and additional held-out correctness cases. Correctness checks enforce the workload's observable utility, allowing implementation changes that preserve it.

Keep candidate compilation and setup outside the timed user operation. Calibrate workload scale, warmups, repetitions, resources, and timeout; then freeze the measurement protocol. Complete verification must fit within 1,200 seconds, and the uncompressed image must not exceed 3 GiB. Target at least three seconds per timed execution; shorter operations require a clear signal and repeated CV at most 10%.

Check fresh-image correctness and flakiness. Diagnose variance before increasing workload size or resources. Enforce the same runtime policy for baseline and candidate. Run a baseline and one bounded candidate-agent attempt, retaining the trajectory and raw measurements. One correct, repeatable gain is sufficient; an incorrect or no-gain candidate does not reject the task if its trajectory demonstrates understanding and a reasonable optimization attempt.

Human review: Reliability. Present fresh-build correctness, raw measurements, dispersion, verification time, resources, and the candidate trajectory. Ask whether equivalent runs support the same verdict and measurement.

## Stage 3: Validate Integrity

Audit agreement between instructions and verifier checks, including writable paths, workload copies, expected outputs, caches, and reward inputs. Run explicit red-team tests for leaked-data access, verifier modification, incomplete work, hard-coded answers, workload-specific precomputation, frozen runtime-policy overrides, and reward-path bypasses.

Count an attack only when a materially distinct invalid implementation reaches the final verifier. After hardening, replay the attack and record its verdict. Reopen Reliability if changes affect the workload, measurement, resources, or verifier execution.

Human review: Integrity. Present the boundary audit and red-team results. Ask whether an invalid or shortcut implementation can receive a valid score. Verifier ownership requires an independent immutable workload copy, not secrecy of the visible workload.

## Stage 4: Finalize

Summarize review decisions and limitations, remove construction artifacts from the release package, and run final checks. Confirm that the package contains all required inputs and no credentials, private evidence, or hidden answers.

Ask the human to ACCEPT AND PUBLISH, REVISE, or REJECT. Publish the pinned image and verified task package only after acceptance; preserve evidence for rejected candidates.
\end{tcblisting}

\subsection{Self-Review Skill}
\label{app:boxbox-review-skill}

% Source: dev/recursive-autoresearch at d741589cd15aa3f75c0acda412380c6ed43f5226,
% perfbench/run/agents/recurnus/runtime/recurnus/skills/selfrewrite/SKILL.md.
% The domain-specific causal-chain sentence, cost-model/checklist/measurement
% block, and workload/cost clause in the Early-stage item are omitted.
% The heading is renamed Self-Review; YAML metadata is omitted and prose rewrapped.
% Research-search terminology is normalized to task-solving terminology.

The following listing presents the default review skill with performance-optimization-specific guidance omitted. ``Target Agent'' denotes the task-solving agent.

\begin{tcblisting}{paperprompt,
  title={Self-Review Skill},
  title after break={Self-Review Skill \normalfont\itshape (continued)}
}
# Self-Review

Act as the macro-level controller of the Target Agent's task-solving process. Review the trajectory to diagnose the task-solving process, then retain the current harness or select a focused-feedback or reusable-ability adaptation that improves subsequent Target progress within the remaining shared budget.

## Review the task-solving process

Treat each review as one iteration in a continuing self-review loop. At the start of every review, examine the accumulated trajectory comprehensively before selecting an intervention.

Use that broad diagnosis to choose a focused intervention that is supported by the evidence and likely to materially improve subsequent Target progress. Include only the feedback or reusable capabilities justified by the diagnosed problems in task execution and the evidence needed to evaluate their effect.

Never be satisfied with the current optimization state while task budget remains. Treat accepted improvements as intermediate evidence and compare the remaining opportunities by their expected contribution.

### Diagnose the task-solving process

Treat the Target Agent's framing, causal model, plans, explanations, and claims as hypotheses, not premises. Use recorded actions, observations, and measurements to identify all material unsupported links, distinguish evidence from inference, and identify plausible explanations not ruled out. Use the complete diagnosis to keep the intervention focused on the related gaps that most materially constrain progress. Treat claims that an experiment completed, improved performance, or excluded a direction as unverified until supported by the trajectory. Explain how each unsupported link affects task-execution cost, experiment selection, or progress.

Critique the Target Agent's choice of actions, not just candidate outcomes. Match each conclusion to the scope the experiments actually tested. Do not generalize beyond the tested workload, configuration, mechanism variant, or measurement, and do not ignore what the evidence establishes within that scope. Distinguish direct findings from broader hypotheses, and identify unsupported assumptions or omitted dimensions that constrained the search.

After selecting the intervention focus, review two aspects of macro-level control as ways to advance it:

1. **Task-execution cost.** Examine repeated Target Agent work, which dependencies and evidence requirements have stabilized, and where intermediate model judgment remains valuable. Look for opportunities to reduce, fuse, automate, batch, or reuse work while preserving decision-relevant evidence.
2. **Budget allocation.** Evaluate exploration breadth and exploitation depth by comparing the causal support, remaining leverage, uncertainty, and expected contribution of validated mechanisms and untested hypotheses.

## Decide the intervention

After completing the diagnosis, choose among retention, a focused-feedback adaptation, and a reusable-ability adaptation according to their expected contribution to target progress over the remaining budget. Retention means no new Target-facing guidance or other harness change.

Use focused feedback for a task-specific concern, unsupported judgment, omission, or unresolved evidence. Identify the concrete observation, the gap, and why it matters. **Be careful about your own claims:** an evidence gap does not establish an alternative explanation or direction; do not turn preferences into concerns or next-step menus. State only the concern and its evidentiary basis; do not require or recommend what the Target Agent should do next.

Activate a reusable ability when it directly supports the diagnosis and the trajectory supports repeated use. Also consider a harness adaptation when the Target Agent repeatedly makes the same consequential error and the evidence supports a general way to prevent it; otherwise use focused feedback. For a stable repeated action sequence, fuse or automate the mechanical work while preserving model judgment, decision-relevant evidence, meaningful failure boundaries, and a fallback to the unfused workflow. Validate the ability through its actual Target-facing path and use feedback to explain how to use it and its constraints.

Condition each intervention on the current stage of task execution, inferred from the trajectory rather than elapsed time:

- **Early:** examine plausible competing hypotheses.
- **Developing:** reduce demonstrated repeated work and balance exploration with supported exploitation.
- **Saturated:** identify which tested hypothesis families have low remaining value, and flag any claim that the broader search is exhausted unless the trajectory supports that scope.

Use the observed downstream effects of prior adaptations when deciding what the current stage needs.
\end{tcblisting}